# A Comparison of Whole Slide Image Analysis and Diagnostic Field Selection in Pathology AI under Finite Resources

Tatsuaki Tsuruyama†, M.D., Ph.D.

Department of Drug Discovery Medicine, Graduate School of Medicine, Kyoto University

Yoshida-Konoe-cho, Sakyo-ku, Kyoto 606-8501, Japan

Running title: Diagnostic Field Selection under Finite Resources

†Corresponding author: email: tsuruyam@kuhp.kyoto-u.ac.jp

**Abstract**

An important question in pathology AI is how to allocate limited analytical resources between broad coverage of a whole slide image (WSI) and detailed analysis of selected regions. We compared a minimal WSI-AI model designed for broad search with Diagnostic Field Selection AI (DFS-AI), which uses a coarse overview to select diagnostically relevant fields for detailed analysis. Three minimal models introduced localization information, contextual discrimination, and spatial structure in sequence. In Model I, WSI-AI was favored when coarse localization was uninformative or costly. When coarse observation provided useful ranking information, DFS-AI was favored over an intermediate range of resources; WSI-AI again became preferable when resources allowed nearly complete fine observation. In Model II, background heterogeneity generated distractors. In a WSI control given the same local context as DFS-AI, most of the original DFS advantage was explained by local background correction, while coarse candidate selection provided a smaller additional benefit. In Model III, larger contiguous lesions were easier for WSI-AI to discover but more difficult to characterize completely with a finite budget for fine observation. The DFS advantage for complete characterization persisted when the comparator used the same field sized unit of fine analysis and when success was defined symmetrically as 100%, 90%, or 80% lesion unit coverage. Together, the models support a framework in which the preferred strategy changes with available resources and depends on the information available for selection, its acquisition cost, lesion structure, and the diagnostic endpoint. This framework clarifies when limited pathology AI resources should be devoted to broad WSI coverage and when they should be concentrated on selected Diagnostic Fields.

## Introduction

Artificial intelligence (AI) in digital pathology has developed from multiple instance learning (MIL) to long sequence Transformers and whole slide foundation models. These methods have greatly improved the ability to extract diagnostic information from an entire whole slide image (WSI) [5–8]. However, having access to the whole WSI is not the same as keeping all information needed for diagnosis in the final representation when resources are limited. In actual WSI-AI systems, the WSI is divided into many tiles. Each tile is compressed into an embedding. The embeddings are then selected and combined into a representation of limited length.

Information needed for pathology diagnosis is also not spread evenly across a WSI. Histologic type is judged from a combination of cell morphology and tissue architecture. Invasion is judged from the relationship between the lesion and surrounding tissue. Inflammation is judged from a combination of cell type, tissue compartment, distribution, tissue injury, and time phase. These tasks often require observation of several fields at several magnifications. The observer may first understand the overall architecture at low magnification, then select borders or tissue compartments at medium magnification, and finally confirm cell features at high magnification. In contrast, when the main task is simply to find a rare target, such as a small lymph node metastasis, broad coverage is important.

For pathology AI, the important question is therefore not only which AI architecture to use. It is also what image information should be given to the AI when analysis resources are limited. This study compares WSI-AI and DFS-AI as two ways of allocating image analysis resources. In WSI-AI, the whole WSI is the target of analysis and the system selects, compresses, and aggregates image information. In the limited resource models used here, WSI-AI denotes a coverage oriented minimal comparator that does not use observer field selection in advance and instead distributes available fine observation resources across a broad search. This comparator is not intended to represent every modern WSI architecture; real WSI systems can themselves use local context, multiscale aggregation, learned attention, or selective sampling. In DFS-AI, an observer selects the image information needed for diagnosis as Diagnostic Fields, and the AI analyzes the selected information. A Diagnostic Field is not a single rectangular ROI. It can include several regions, magnifications, and comparison relationships chosen for the diagnostic purpose. For example, diagnosis of histologic type may require low magnification tissue architecture and high magnification cell features; evaluation of invasion may require the lesion, invasive front, and surrounding tissue; and evaluation of inflammation may require relationships among cell features, tissue compartments, and distribution. We do not limit the source of information used to select a Diagnostic Field. DFS-AI includes both direct observer selection and workflows in which the observer reviews candidate regions suggested by WSI-AI and makes the final choice of fields. The observer remains outside both methods as the final diagnostician, and AI output is treated as diagnostic support rather than independent ground truth.

Recent approaches in computational pathology increasingly use selective sampling, hierarchical processing, or learned viewing policies to reduce the amount of whole slide information that must be processed at high resolution [10, 13, 16]. These studies demonstrate the practical value of selective computation, but they do not by themselves determine when selective analysis should be preferred to broad whole slide coverage. The unresolved problem is one of resource allocation. When a fixed analysis budget can be spent either on broader fine resolution coverage or on an overview acquired at relatively low cost followed by focused fine analysis, which strategy is expected to preserve more diagnostically relevant information? The contribution of this study is not a new image selection architecture, but a framework for determining how the preferred strategy changes with available resources. Across three minimal models, the information available for selection progresses from coarse localization in Model I, to contextual discrimination in a heterogeneous background in Model II, and to structural aggregation of a spatially continuous lesion in Model III. The main quantity of interest is the crossover boundary at which the information gained through selection outweighs, or is outweighed by, the costs in observation resources and coverage.

## Results

### Under limited capacity, information from regions with low diagnostic relevance can compete for representational capacity

Let $S$ be the image structures needed for diagnosis. Let $N$ be the remaining image information that provides little additional information for the diagnostic task once $S$ is known. We write the WSI as $X = (S, N)$. In an idealized case, assume

$$I(Y;N \mid S) = 0 \quad (1)$$

This means that once $S$ is known, adding $N$ does not reduce uncertainty about diagnosis $Y$. Then

$$I(Y;S,N) = I(Y;S) \quad (2)$$

Thus, even though $N$ contains image information, it adds no further diagnostic information. Let $Z_W$ be the finite length internal representation of WSI-AI and $Z_D$ the internal representation of DFS-AI. Let $C$ be the upper limit on the amount of information that either representation can retain. Let the amount of low relevance information that remains in the WSI-AI representation be

$$c_N = I(N;Z_W \mid S) \quad (3)$$

To make the restricted capacity argument explicit, we assume that the WSI representation $Z_W$ is generated only from $X = (S, N)$, that the representation obeys a finite capacity constraint $I(X; Z_W) \leq C$, and that the nuisance information retained in $Z_W$ shares the same finite capacity with diagnostically relevant information. In this accounting model, $I(Y; Z_W) + c_N \leq C$. Together with $I(Y; N \mid S) = 0$ and the corresponding data processing bound $I(Y; Z_W) \leq I(Y; S)$, this yields the upper bound below. This is a stylized resource allocation assumption used to isolate capacity competition; it is not a universal information inequality for arbitrary encoders.

With this definition, the restricted model gives the following upper bound on the diagnostic information that WSI-AI can retain:

$$I(Y;Z_W) \leq \min\{I(Y;S),\ C - c_N\} \quad (4)$$

Equation (4) gives this upper bound. As $c_N$ becomes larger, less of the finite capacity $C$ can be used for diagnostically relevant structure. Here, $c_N$ is not the number of bytes in GPU memory. It is a theoretical quantity that represents the amount of low relevance variation remaining in the finite representation.

**Core condition for an advantage of DFS**

Let the diagnostic information that could ideally be retained under capacity $C$ be

$$I_0(C) = \min\{I(Y;S), C\} \quad (5)$$

We define the total loss as the information lost because the observer did not select a needed field, because a needed magnification or comparison was missing, or because the selected fields were compressed into a finite representation.

$$d_D(C) = I_0(C) - I(Y;Z_D) \quad (6)$$

We define this quantity in Equation (6). In contrast, let the loss caused by the WSI-AI upper bound above be

$$g_W(C) = I_0(C) - \min\{I(Y;S), C - c_N\} \quad (7)$$

With these definitions,

$$d_D(C) < g_W(C) \quad (8)$$

If the condition in Equation (8) holds, it is sufficient for the DFS-AI internal representation to retain more diagnostic information than the WSI-AI representation. In the range where diagnostically relevant information alone uses all of capacity $C$, this condition becomes

$$d_D(C) < c_N \quad (9)$$

Equation (9) has a simple interpretation. DFS-AI may retain more diagnostic information if the diagnostic information lost through observer field selection is smaller than the diagnostic information lost in WSI-AI because low relevance regions compete for storage, selection, and aggregation. Conversely, WSI-AI is favored when the observer misses too much important information during selection.

**Complementary information across multiple Diagnostic Fields**

Let the two diagnostic fields be $A$ and $B$. For example, let $A$ be low magnification tissue architecture and $B$ be high magnification cell features. If

$$I(Y; B \mid A) > 0 \tag{10}$$

and

$$I(Y; A \mid B) > 0 \tag{11}$$

then each field carries diagnostic information that is not fully contained in the other. This establishes complementarity across fields, not necessarily formal synergistic information. DFS should therefore not be designed as a single ROI cropped only from the lesion; it should select several fields while preserving the relationships needed for diagnosis.

## 1 Model I: Search for a single rare lesion under finite resources using WSI and DFS

Model I first leaves aside the morphologic complexity of pathology images and isolates only the difference in limited observation resources and observation strategy. We place $M$ observation points on a square grid. Only one point is a lesion, and all other points are normal. In the figures, the lesion is always red and normal points are always blue. The meaning and color of the lesion do not change. Only its coordinate $L$ is moved across all grid points. In the representative calculations below, we use a 7×7 grid, so $M$ = 49, and assume that lesion location follows the uniform distribution $P(L = i) = 1/M$. Therefore, no coordinate has a diagnostic prior.

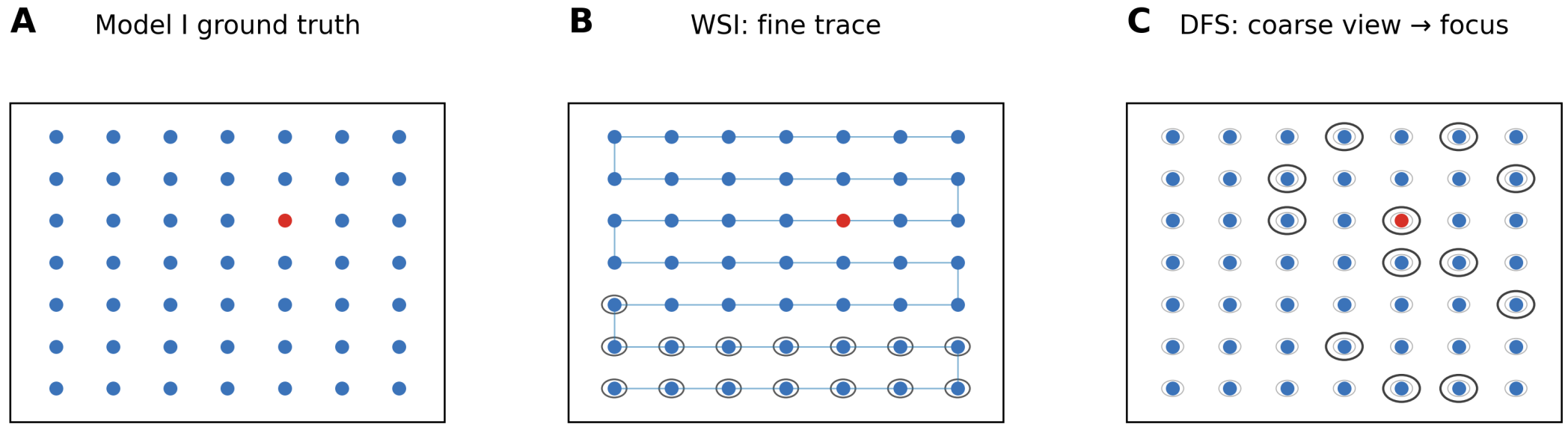

**Figure 1.** Observation strategies in Model I. A: Ground truth with one lesion point (red) and normal points (blue). B: WSI-AI traces the entire area with fine observation from the start. C: DFS-AI first makes a low-cost coarse view of the entire area, ranks candidate points by the coarse score, and then performs fine analysis only on selected points. In the calculation, the red point is moved through all 49 coordinates.

Let the total observation budget be $B$, and let the cost of fine observation at one point be $c_f$. Model I assumes that a lesion is detected perfectly if it is present at a point that receives fine observation. We do not include error from the fine detector itself. WSI-AI does not perform coarse selection in advance. It starts with fine observation and applies it point by point. The number of points that can be observed, $k_W$, and the detection probability $P_{WSI}$ averaged over all lesion locations are

$$k_W = \min\left\{M, \mathrm{floor}\left(\frac{B}{c_f}\right)\right\}, \qquad P_{\mathrm{WSI}} = \frac{k_W}{M}. \tag{12}$$

With a fixed trace order, success or failure at each coordinate depends on the order. However, if the lesion is moved uniformly across all $M$ coordinates and the result is averaged, the detection probability is $k_W/M$. It does not depend on the coordinate labels or on the geometric way the trace is drawn.

### 1.1. DFS-AI: Coarse localization followed by focused fine analysis

In DFS-AI, the observer first views all $M$ points coarsely at low cost. Let the coarse observation cost per point be $c$. The fixed cost of the whole coarse view is $Mc$. After the coarse view, the number of points that can receive fine analysis, $m_D$, is

$$m_D = \min\left\{M, \max\left[0, \mathrm{floor}\left(\frac{B - Mc}{c_f}\right)\right]\right\}. \tag{13}$$

We represent how much ranking information the coarse view has about lesion location by the separation $\delta$ between the coarse scores of normal and lesion points. Let the score of a normal point be $Z_0$ and the score of the lesion point be $Z_1$, with

$$Z_0 \sim \mathcal{N}(0,1), \qquad Z_1 \sim \mathcal{N}(\delta, 1). \tag{14}$$

When $\delta = 0$, the coarse view contains no information about lesion location. As $\delta$ increases, the coarse view can rank lesion candidates closer to the top. DFS-AI selects the top $m_D$ points by coarse score and performs fine analysis only on those points. The process called coarse overview → focus here is not a third method. It is the observation process of DFS itself.

### 1.2. Exact detection probability for DFS

Suppose the coarse score of the lesion point is $z$. The probability that one normal point has a score higher than $z$ is $1-\Phi(z)$, where $\Phi$ is the standard normal cumulative distribution function and $\varphi$ is its density. For the lesion point to be among the top $m_D$ points, at most $m_D-1$ of the $M-1$ normal points may have scores above $z$. Therefore, the detection probability averaged over coordinates of DFS can be calculated directly by the following integral, without Monte Carlo simulation.

$$P_{\mathrm{DFS}} = \int_{-\infty}^{\infty} \phi\,(z-\delta) \sum_{j=0}^{m_D-1} \binom{M-1}{j} [1-\Phi(z)]^j [\Phi(z)]^{M-1-j}\, dz. \tag{15}$$

Equation (15) directly shows the tradeoff between the value of ranking information from the coarse view and the fixed cost $M\,c$ needed to obtain that information. DFS is favored over WSI when $P_{DFS} > P_{WSI}$. The boundary can be defined by the critical discriminability $\delta^*$ that satisfies $P_{DFS} = P_{WSI}$.

### 1.3. Representative calculation

Below, $M = 49$, the fine observation cost is $c_f = 1$, and the coarse view cost is $c = 0.05$. The fixed cost for DFS to view all 49 points coarsely is 2.45. We use $\delta = 0.5$ as a representative coarse view discriminability. At $B = 15$, WSI-AI can observe 15 points finely, so $P_{WSI} = 15/49 = 0.3061$. Using Equation (13), DFS-AI can observe 12 points finely after the coarse view, and the exact integral in Equation (15) gives $P_{DFS} = 0.4207$. Thus, under this condition, the absolute advantage of DFS is 0.1146. At $B = 15$, the critical coarse view discriminability at which the two methods are equal is $\delta^* = 0.1873$. Even fairly weak coarse localization information can therefore repay the cost of the coarse view by focusing fine observation under intermediate limited resources.

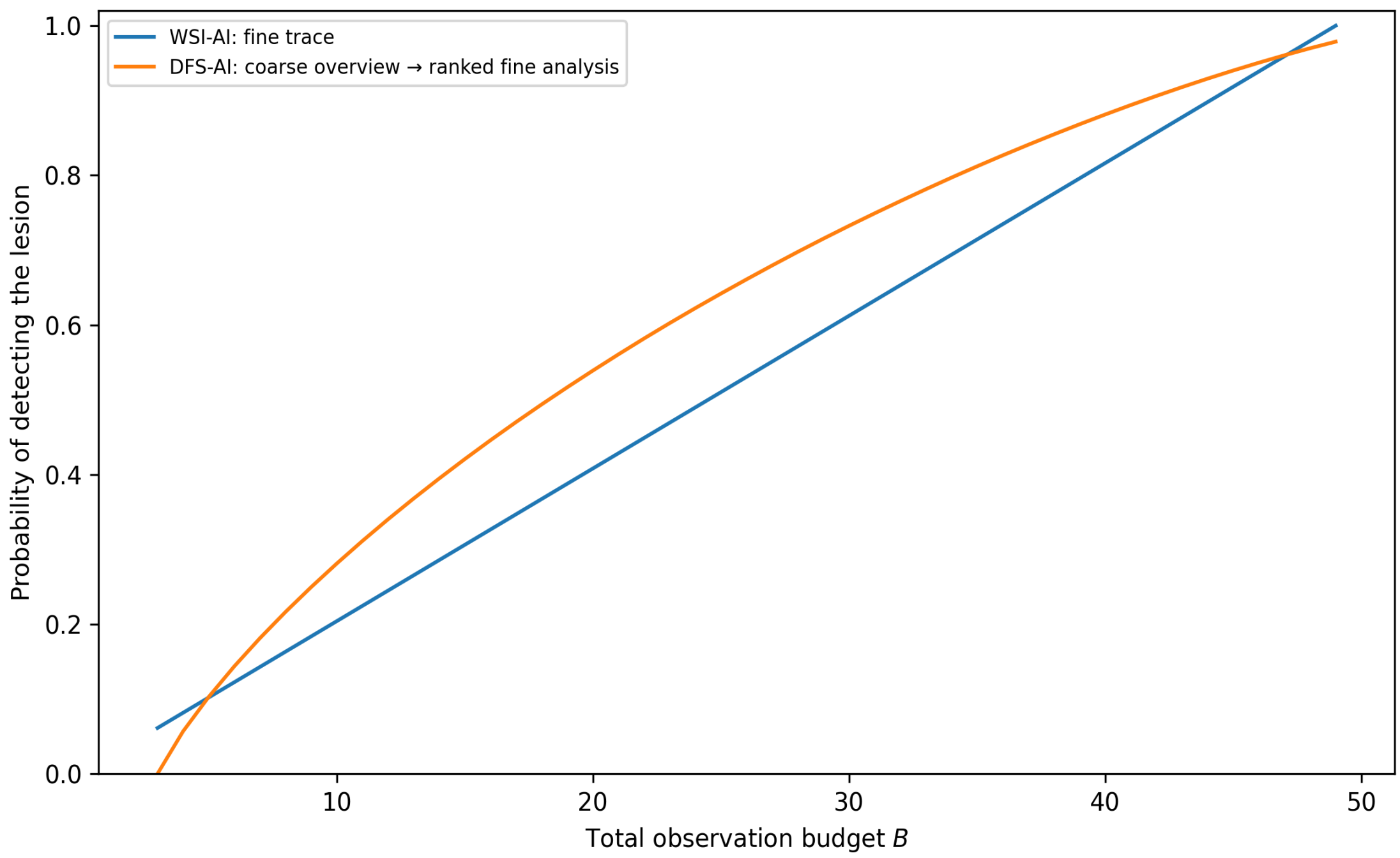

**Figure 2.** Total observation budget $B$ and detection probability in Model I. $\delta = 0.5$ and $c = 0.05$. At very low resources, the fixed cost of the coarse view dominates and WSI is favored. At intermediate resources, DFS is favored. At high resources, where almost the entire area can be observed finely, WSI is favored again.

### 1.4. Resource dependent strategy crossover

The main result of Model I is that neither method is always best. The preferred strategy changes with the total resources and the amount of information available from the coarse view. When $\delta = 0$, the coarse view gives no information at all about lesion location, and there is no resource range in which DFS is favored. When $\delta = 0.2$, DFS is favored only for $B = 14$–$38$. When $\delta = 0.5$, DFS is favored for $B = 5$–$47$. When $\delta = 1.0$ or $\delta = 2.0$, the range expands to $B = 4$–$48$. Thus, as the coarse view provides more information, the limited resource range in which DFS is favored becomes wider.

Low resources: WSI → Intermediate resources: DFS → High resources: WSI

At low resources, DFS has a largely fixed cost for the coarse view of the entire area. Too few resources remain for fine observation, so WSI is favored. In the practical intermediate resource range, the value of using ranking information from the coarse view to focus fine observation on likely candidates is greater than the fixed cost, so DFS is favored. In an ideal high resource range, WSI can observe almost every point finely. The value of saving resources by selection then disappears, and WSI, which has no coarse view cost, is favored again.

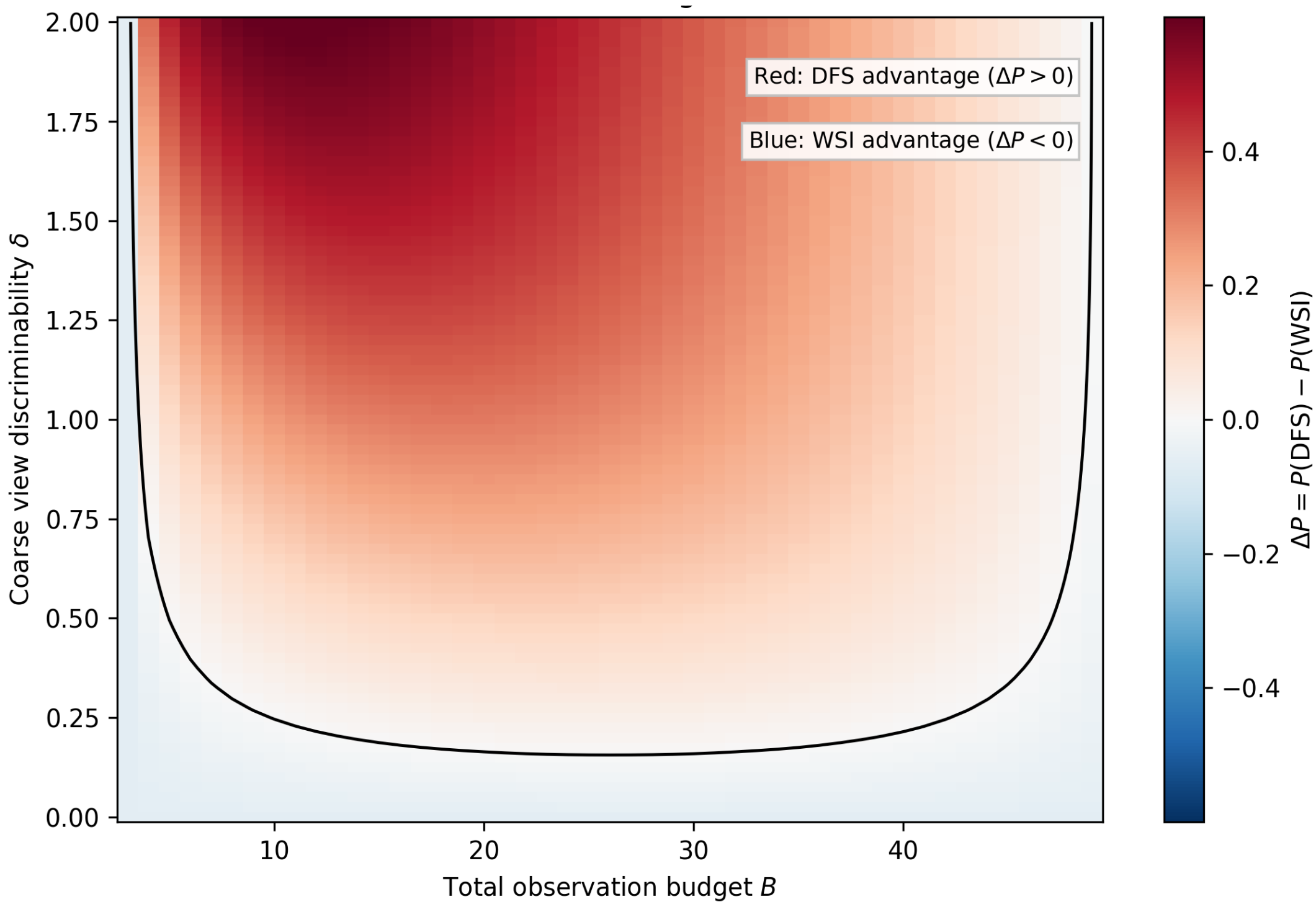

**Figure 3.** Advantage of DFS as a function of total observation budget $B$ and coarse view discriminability $\delta$. Color represents $\Delta P = P(\mathrm{DFS}) - P(\mathrm{WSI})$: red indicates $\Delta P > 0$ (DFS advantage), blue indicates $\Delta P < 0$ (WSI advantage), and the black zero contour is the boundary between the two methods.

### 1.5. Sensitivity to the cost of coarse observation

The gain of DFS does not assume that the coarse view is free. As $c$ increases, the fixed cost $M\,c$ increases. The number of points $m_D$ that can receive fine analysis at the same $B$ therefore decreases, and the range in which DFS has an advantage becomes smaller. Thus, the advantage of DFS does not come only from the fact that an observer selects the fields. It depends on the tradeoff between cost and information: how much location information a low cost coarse observation can provide. The fixed budget section in Figure 4B shows the same tradeoff directly: at $B = 30$, increasing the coarse view cost shifts $\Delta P$ downward, while stronger coarse view information $\delta$ preserves a DFS advantage over a wider cost range.

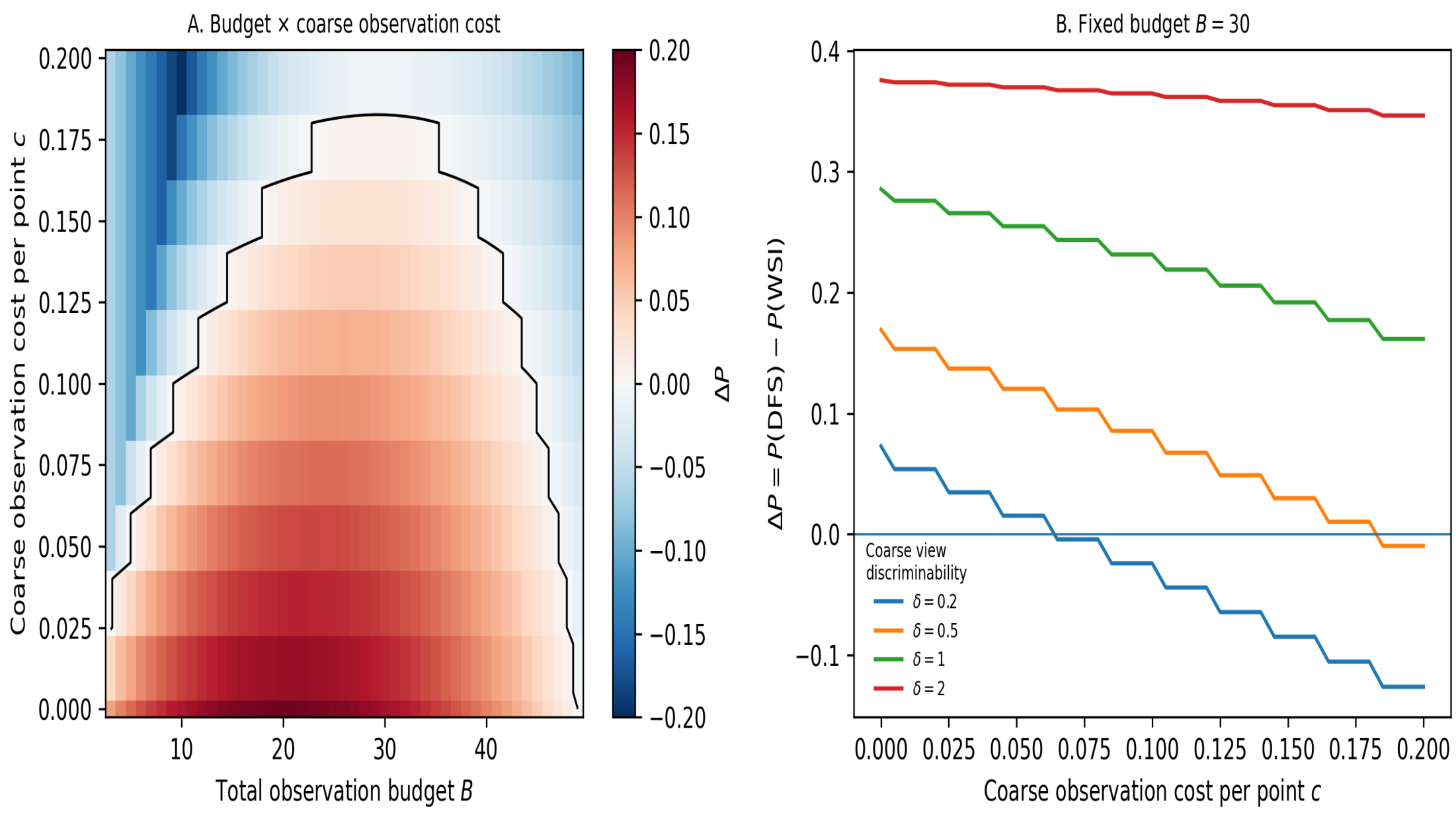

**Figure 4.** Sensitivity of Model I to coarse view cost $c$. A: $\Delta P$ = $P$(DFS) − $P$(WSI) as a function of total observation budget $B$ and coarse view cost $c$ with $\delta$ = 0.5; red indicates a DFS advantage and blue indicates a WSI advantage, and the black contour is $\Delta P$ = 0. B: fixed budget profiles at $B$ = 30 for $\delta$ = 0.2, 0.5, 1.0, and 2.0. Increasing $c$ reduces the DFS advantage, whereas stronger coarse view information makes the advantage more resistant to coarse view cost.

### 1.6 Robustness to lesion coordinates and conclusion of Model I

Because lesion location $L$ is moved uniformly across all $M$ coordinates in this model, the final result does not depend on where the red point is drawn in the figure. In DFS, the coordinates are exchangeable, so under the same $\delta$ the lesion has the same rank distribution at every coordinate. With the fixed WSI trace, the time at which each coordinate is found depends on the order. However, the mean success rate across all coordinates is $k_W/M$, which is unchanged by relabeling the coordinates. The advantage shown by Model I therefore comes from the rule used to distribute limited resources, not from a particular choice of location.

These results show that even in the smallest problem of one rare lesion, the best observation strategy changes with the amount of resources and the information available from the coarse view. WSI is favored when the coarse view has no useful information or is too costly. DFS is favored in an intermediate range of limited resources when a low cost coarse view gives enough ranking information about lesion location and fine observation of the entire area is not possible. When resources become large enough for WSI to observe the entire area finely, WSI is favored again. The advantage of DFS is therefore not simply that it observes a smaller area. Its advantage is that location information from the coarse observation can be used to move expensive fine analysis resources toward more useful locations. Model I is a minimal model with perfectly fine observation, one lesion, and a uniform background. Background heterogeneity, distractors, local contextual information, and structural dependence are introduced in the subsequent models.

## 2 Model II: Detection of an infectious organism in a heterogeneous background

Background tissue in pathology specimens is not always uniform. Model II is intended primarily to represent searches for infectious organisms, with analogous settings such as leukemic cells in blood also in mind. Normal tissue can vary in color, cell density, structure, and other features, and some regions can therefore resemble a lesion. In this setting, simply searching for a visually conspicuous region is insufficient; background variation must be distinguished from findings that are specific to the target.

The task considered here is detection of a red signal from an infectious organism in an environment in which background color varies by location. When background heterogeneity is small, the organism can be identified relatively easily from absolute redness. As background heterogeneity increases, however, noninfected regions can also become strongly red and act as distractors.

In DFS, candidate selection can use comparison with the surrounding tissue rather than the redness of the target alone. Reference to the local background can remove part of the color variation caused by background tissue and make the organism specific signal easier to distinguish. WSI-AI is modeled as a strategy that directly performs detailed observation of many fields under a finite observation budget. Model II examines numerically how background heterogeneity changes the relative detection performance of DFS and WSI-AI.

## 2.1 Modeling the background and infectious organism

The observation area contains $M$ candidate points, and one of them contains the infectious organism.

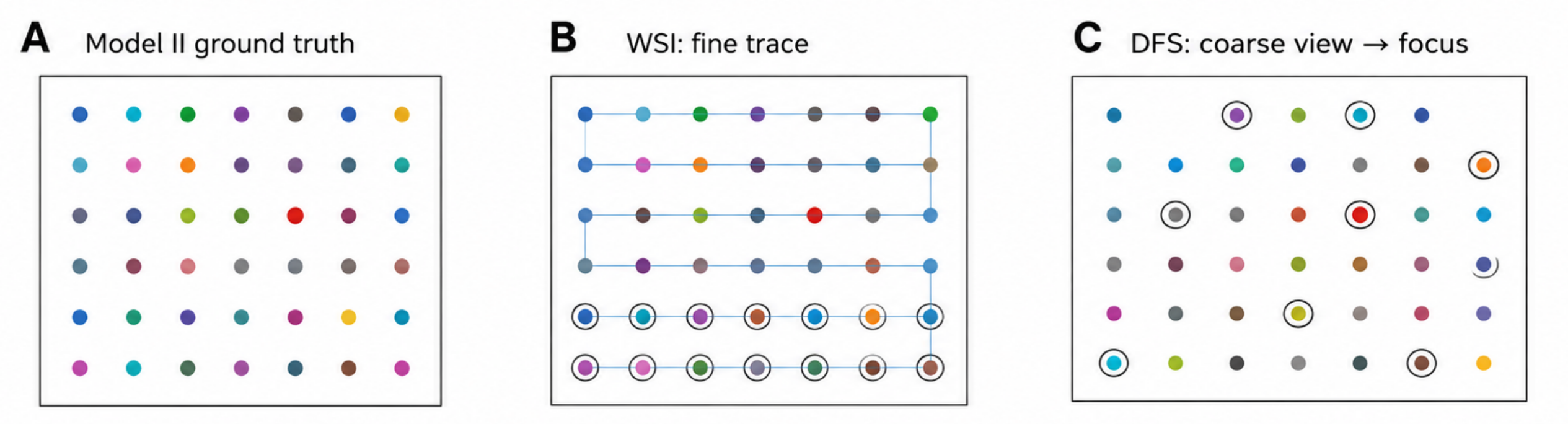

**Figure 5.** Observation strategies in Model II. A: Ground truth. Each point has a different color, representing background heterogeneity, and one red target represents the infectious organism. B: WSI-AI distributes the available fine observation resources across a broad search; nonlesion points that resemble the target act as distractors. C: DFS-AI first uses a coarse overview and local background context to rank candidates, then performs focused fine observation on selected points. Circled points indicate targets of detailed observation.

Let $i$ denote a candidate point. Set $I_i = 1$ when candidate point $i$ contains the infectious organism and $I_i = 0$ when it does not. Define the latent redness at the center of each candidate point, $x_i$, as

$$x_i = b_i + \eta_i + \Delta I_i \tag{16}$$

Here, $b_i$ is the background color component at candidate point $i$, $\eta_i$ is a small color variation specific to that site, and $\Delta$ is the size of the red signal specific to the infectious organism. The background component and the site specific variation are modeled as

$$b_i \; N(0, {\sigma_b}^2), \eta_i \; N\left(0, {\sigma_\eta}^2\right) \tag{17}$$

Here, $\sigma_b$ represents background heterogeneity. When $\sigma_b$ is small, the background colors are fairly uniform. As $\sigma_b$ increases, color differences among candidate points become larger. Even in a noninfected region, a large positive value of $b_i$ can produce a red color similar to that of the infectious organism. We treat such a region as a distractor. An increase in $\sigma_b$ therefore does not simply mean more observation noise. It means that the sample itself contains more candidates that resemble the infectious organism.

**2.2 Local background**

In real tissue observation, the color of a region is often judged not only by its absolute value but also by its difference from surrounding tissue. Model II therefore introduces a local background component $b_i^{\text{local}}$ around candidate point $i$. We assume that the central background $b_i$ and the local background $b_i^{\text{local}}$ are correlated. Let this correlation be $\rho_l$, with $0< \rho_l<1$. More specifically, using independent standard normal variables $u_i$ and $v_i$,

$$b_i \; = \; \sigma_b u_i \qquad (18)$$

$$b_i^{\text{local}} \; = \; \sigma_b \rho_l u_i + \sqrt{(1-\rho_l{}^2)} v_i \qquad (19)$$

With these definitions, $b_i$ and $b_i^{\text{local}}$ have the same variance $\sigma_b{}^2$, and their correlation coefficient is $\rho_l$. When $\rho_l$ is close to 1, the background color at the center and around it is very similar. As $\rho_l$ becomes smaller, the local background itself becomes more variable, and it becomes harder to estimate the central background from the surrounding area. DFS evaluates how likely a candidate is to contain the organism by using the difference between the center and the local background, rather than the absolute redness of the candidate. The DFS score in fine observation is

$$z_i^{(D,f)} \; = \; \left(x_i + \varepsilon_i^f\right) - \left(b_i^{\text{local}} + \varepsilon_i^{(local,f)}\right) \qquad (20)$$

The fine observation error terms were sampled independently as $\varepsilon_i^f \sim N(0, \sigma_f^2)$ for the candidate center and $\varepsilon_i^{\text{local,f}} \sim N(0, \sigma_{\text{local,f}}^2)$ for the local background. All observation error terms were generated independently across candidate points and independently of the latent background components.

The local background subtraction in Equation (20) cancels part of the background color component shared by the center and the surrounding area. However, because $\rho_l<1$, background variation is not removed completely.

In contrast, the WSI-AI comparator in Model II does not explicitly correct for the local background. It uses the absolute redness of candidates seen with fine observation as the basic information for identifying the organism:

$$z_i^{WSI} \; = \; x_i + \varepsilon_i^f \qquad (21)$$

This comparison does not mean that WSI-AI in general cannot use local context. It is a minimal model designed to separate two strategies under limited resources: search of the entire area using absolute features, and selective search using local context.

**2.3 Dilution of the target signal during coarse observation**

When the target organism is small, its red signal may be averaged with the surrounding background at low magnification or in a coarse observation. It may therefore appear weaker than in fine observation. We set the organism signal in coarse observation to $\kappa\Delta$, where $0 < \kappa < 1$. Here, $\kappa$ represents the visibility of the organism signal during coarse observation. We define the latent central value during coarse observation as

$$x_i^c \; = \; b_i + \eta_i + \kappa\Delta I_i \qquad (22)$$

Using this latent value, we define the DFS coarse score, including error in estimating the local background, as

$$z_i^c \; = \; (x_i^c + \varepsilon_i^c) - \left(b_i^{\text{local}} + \varepsilon_i^{(local,c)}\right) \qquad (23)$$

The corresponding coarse observation error terms were sampled independently as $\varepsilon_i^c \sim N(0, \sigma_c^2)$ for the candidate center and $\varepsilon_i^{\text{local,c}} \sim N(0, \sigma_{\text{local,c}}^2)$ for the local background.

With this setting, when the background is uniform and the organism is easy to identify, adding a coarse observation in DFS does not necessarily provide much extra information. In contrast, when background heterogeneity increases and more distractors appear, candidate selection based on the difference from the local background becomes more valuable.

**2.4 Observation strategies for DFS and WSI-AI**

Let $B$ be the total observation budget, $c$ the cost of coarse observation per candidate point, and $c_f$ the cost of fine observation. In DFS, all $M$ candidate points first receive coarse observation. The fixed cost of this stage is $M\,c$. The number of candidate points that can receive fine observation with the remaining budget, $m_D$, is

$$m_D = floor\left((B - Mc)/c_f\right) \tag{24}$$

DFS ranks the candidate points by the coarse score $z_i^c$ and performs fine observation only on the top $m_D$ points. It then chooses the candidate that looks most like the organism using the score after local background subtraction in fine observation. WSI-AI does not perform coarse observation and uses the total observation budget directly for fine observation. Therefore, the number of candidate points that can receive detailed observation, $k_W$, is

$$k_W = floor\left(B/c_f\right) \tag{25}$$

WSI-AI performs fine observation on $k_W$ points in an order that does not depend on location. Because the organism location is placed uniformly over all candidate points and the results are averaged, this is equivalent to uniformly sampling $k_W$ points. Under the same budget $B$, $m_D < k_W$. DFS therefore observes fewer candidate points in detail than WSI-AI. In return, coarse observation and local background information may allow DFS to focus fine observation on candidates that are more likely to contain the organism. The comparison in Model II is therefore a tradeoff between the number of fields that can be observed and the contextual information available for choosing candidates.

**2.5 Definition of successful detection**

In each trial, detection is counted as successful when the true infectious organism is finally selected as the organism candidate. DFS requires success at two stages. First, the true organism must be included among the top $m_D$ points after ranking by coarse observation. Second, among the candidates that receive fine observation, the organism must have a higher final score than the distractors.

For WSI-AI, success means that the true organism is included among the $k_W$ candidates receiving detailed observation and has a higher organism score than the other observed candidates. Background heterogeneity therefore affects the two strategies in different ways. In WSI-AI, as $\sigma_b$ increases, more distractors with high absolute redness appear, making it harder to distinguish them from the organism. DFS can remove the part of the background variation shared with the local background. However, because the local background does not match the center perfectly, very strong background heterogeneity also reduces the discrimination performance of DFS.

**2.6 Numerical simulation**

We performed Monte Carlo simulations of the model described above and compared the probability of detecting the infectious organism with DFS and WSI-AI. In each trial, one of the $M$ candidate points was set as the organism. For each candidate, we generated a central background $b_i$, a local background $b_i^{\mathrm{local}}$, and site specific variation $\eta_i$. An additional red signal $\Delta$ was added to the organism.

For DFS, we first generated a coarse observation for every candidate point and ranked the candidates by the coarse score after local background subtraction. We then performed fine observation on the top $m_D$ points, where $m_D$ was determined by the available budget. The candidate with the highest final score was judged to be the organism. For WSI-AI, we performed fine observation on $k_W$ points in an order that did not depend on location, where $k_W$ was determined by the total budget. Among the observed points, the candidate with the highest organism score was judged to be the organism.

In the simulations, $M = 50$, $\Delta = 1.0$, $\kappa = 0.10$, $\rho_l = 0.90$, and $\sigma_\eta = 0.15$. The observation noise standard deviations were $\sigma_c = 0.70$ and $\sigma_{\mathrm{local,c}} = 0.40$ in the coarse stage, and $\sigma_f = 0.20$ and $\sigma_{\mathrm{local,f}} = 0.15$ in the fine stage. The coarse and fine observation costs were $c = 0.05$ and $c_f = 1.0$, respectively. The total observation budget $B$ was varied from 3 to 50, and background heterogeneity $\sigma_b$ was varied from 0 to 2.0. The value $\sigma_b = 0.70$ was used for the representative budget response in Figure 6A and for the matched control example in Section 2.10. We performed 20,000 Monte Carlo trials at each grid point.

Let $P_{DFS}$ be the success probability of DFS and $P_{WSI}$ the success probability of WSI-AI. We define their difference as

$$\Delta P = P_{\mathrm{DFS}} - P_{\mathrm{WSI}} \tag{26}$$

$\Delta P>0$ indicates an advantage for DFS, whereas $\Delta P<0$ indicates an advantage for WSI-AI.

**2.7 Sensitivity to background heterogeneity**

Figure 6B shows $\Delta P$ while changing total observation budget $B$ and background heterogeneity $\sigma_b$. In the range of low background heterogeneity, there were conditions favoring WSI-AI over DFS. When the background is fairly uniform, the infectious organism can be identified relatively easily from absolute redness alone. In this setting, using the local background to narrow candidates provides little additional information. DFS also pays a fixed cost to perform coarse observation at every candidate point, which leaves a smaller budget for fine observation. Therefore, under an unrealistically simple background, WSI-AI can have an advantage because it can directly observe more candidate points in detail. Figure 6A separately shows the budget dependence of $P_{\mathrm{DFS}}$ and $P_{\mathrm{WSI}}$ at a representative moderate background heterogeneity of $\sigma_b = 0.70$, making the resource effect directly comparable with Model I.

As $\sigma_b$ increased, the relative performance of DFS improved. Greater variation in background color creates noninfected distractors with high redness. If the organism is identified only by absolute redness, it becomes harder to separate the organism from noninfected regions. DFS uses the difference between the central color and the surrounding local background, so it can remove the part of background variation shared by the two. At moderate background heterogeneity, the combined information gained from local contextual discrimination and focused candidate selection was larger than the cost of coarse observation, and DFS had a higher detection probability than the original minimal WSI-AI comparator. When background heterogeneity became even larger, however, the gain of DFS did not continue to increase. The local background and the central background do not match perfectly, so residual variation remains even after local background subtraction. With an extremely complex background, DFS also has difficulty separating the organism from distractors.

These results show that the relative value of DFS does not increase steadily with background heterogeneity. It changes with the complexity of the background. When the background is simple, the broad coverage of WSI-AI is an advantage. As background variation creates more distractors that resemble the organism, DFS candidate selection using local context becomes more useful. If the background variation becomes extremely large, however, local background correction also reaches its limit in DFS.

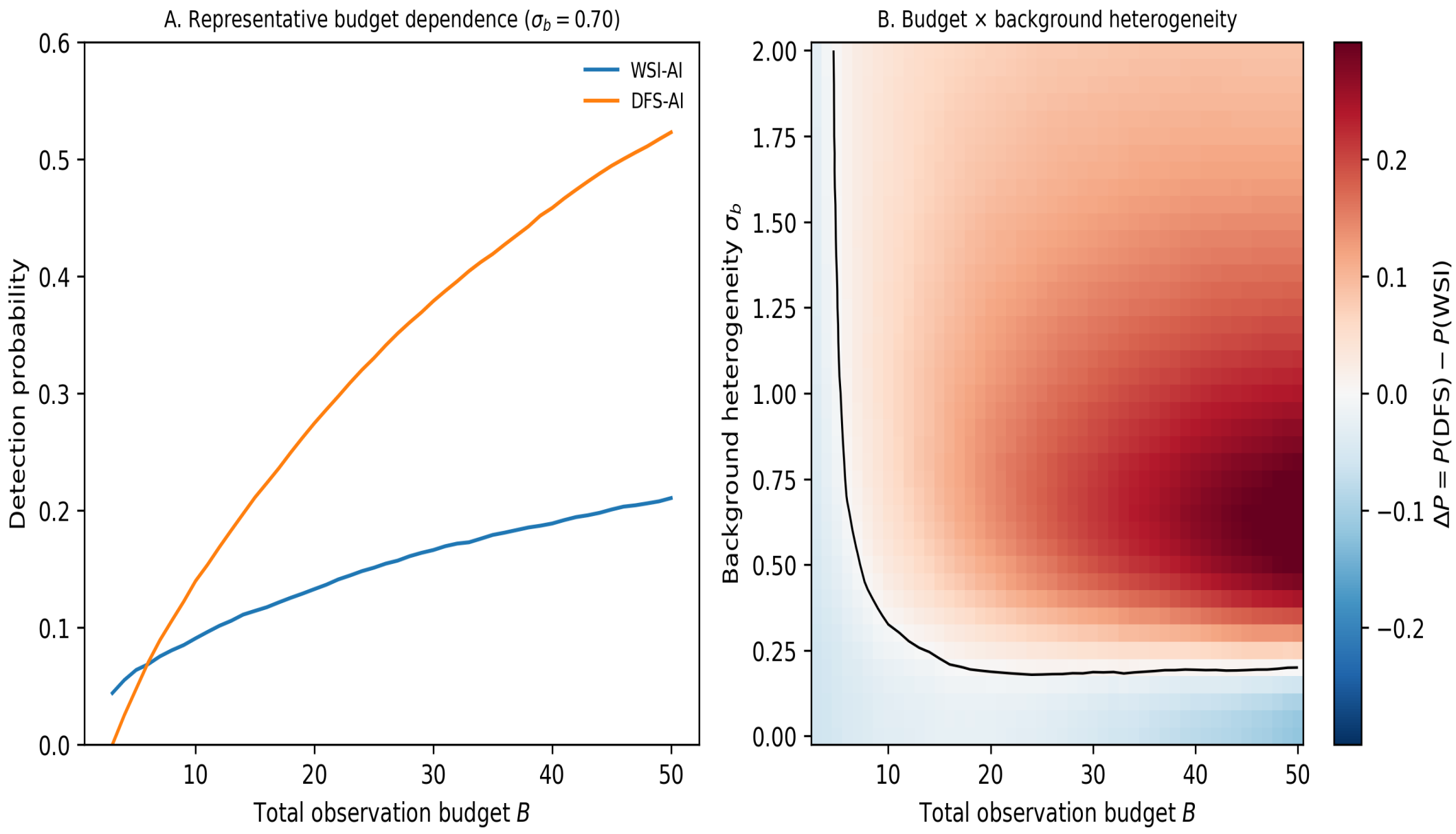

**Figure 6.** Sensitivity of Model II to background heterogeneity $\sigma_b$. Panel B: the horizontal axis is total observation budget $B$, the vertical axis is background heterogeneity $\sigma_b$, and the color shows $\Delta P$. The $\Delta P =$ 0contour is the boundary at which the advantage switches between DFS and WSI-AI. WSI-AI is favored when background heterogeneity is small. As heterogeneity increases, distractors raise the relative value of DFS, and a region of DFS advantage appears at moderate heterogeneity. At still higher heterogeneity, residual variation remains after local background subtraction, so the DFS gain becomes smaller again. Panel A: representative detection probabilities of WSI-AI and DFS-AI as functions of total observation budget $B$ at $\sigma_b$ = 0.70. The blue/red sign convention is therefore identical to Model I.

### 2.8 Sensitivity analysis of key assumptions

To check whether the results of Model II depended only on one parameter setting, we performed sensitivity analyses for three key assumptions: visibility $\kappa$ of the organism signal during coarse observation, correlation $\rho_l$ between the central and local backgrounds, and coarse view cost $c$ per candidate point. We fixed a representative total observation budget at $B = 30$ and used the baseline values for the other parameters. For each condition, we ran Monte Carlo simulations and compared changes in $\Delta P$ across background heterogeneity $\sigma_b$.

#### 2.8.1 Visibility of the organism signal

Here again, $\kappa$ represents how much of the organism specific signal $\Delta$ seen in fine observation remains in coarse observation. When $\kappa$ is small, the organism signal is strongly diluted in the coarse view. As $\kappa$ becomes larger, the organism is easier to recognize already at the coarse observation stage.

For $\kappa = 0.05, 0.10, 0.20$, the basic crossover was preserved: WSI-AI was favored at low background heterogeneity, and DFS gained an advantage as $\sigma_b$ increased. The values of $\sigma_b$ at which the preferred strategy changed were about 0.21, 0.19, and 0.10, respectively. The larger $\kappa$ was, the smaller the background heterogeneity needed before DFS gained an advantage. This is because more information for identifying the organism remains in coarse observation and can be used by DFS to select candidates.

In contrast, when $\kappa = 0.40$, DFS already outperformed WSI-AI when background heterogeneity was almost zero, and there was no clear crossover. This shows that the advantage of WSI-AI in a uniform background is not an unconditional property of Model II. If the organism can already be recognized clearly during coarse observation, selection by DFS can be useful even when the background is simple. Therefore, the pattern of WSI-AI for a uniform background and DFS for a heterogeneous background requires the condition that the organism signal is diluted to some extent during coarse observation.

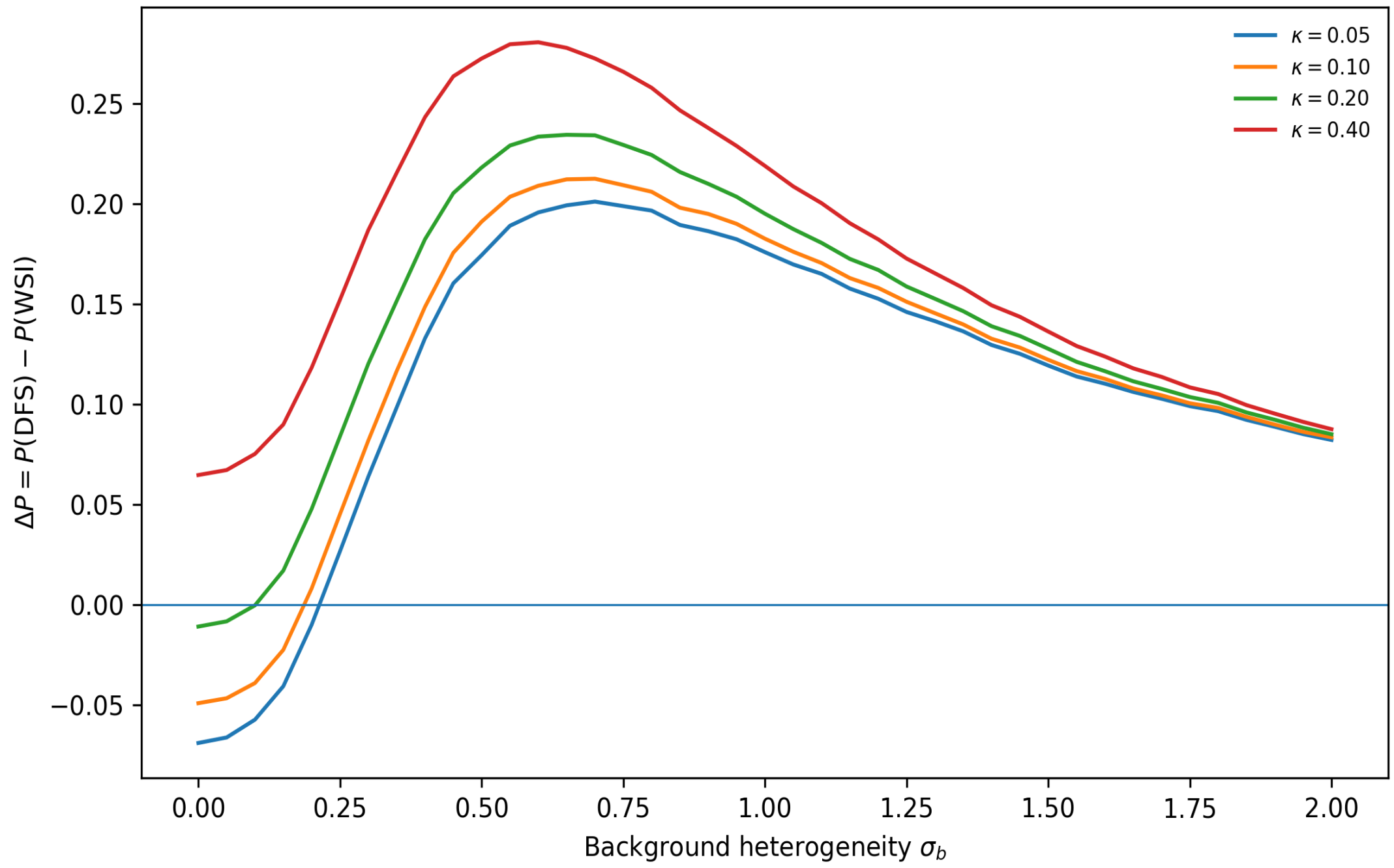

**Figure 7.** Sensitivity to the visibility $\kappa$ of the infectious organism signal during coarse observation. Total observation budget is fixed at $B$ = 30, and $\Delta P$ is shown as a function of background heterogeneity $\sigma_b$. As $\kappa$ increases, more organism information remains in coarse observation, and the shift to a DFS advantage occurs at a smaller $\sigma_b$. When $\kappa$ is large enough, DFS retains an advantage even with a uniform background.

### 2.8.2 Local background correlation

Next, we changed the correlation $\rho_l$ between the background color at the center and in the surrounding local area. A higher correlation allows the central background component to be estimated more accurately from the surrounding background, making local background subtraction more effective.

For $\rho_l$ = 0.70, 0.80, 0.90, and 0.97, the same basic pattern was seen. WSI-AI was favored when background heterogeneity was small, and DFS gained an advantage as $\sigma_b$ increased. The change to a DFS advantage occurred at about $\sigma_b$ = 0.17–0.24. The presence of the crossover was therefore not limited to one specific $\rho_l$.

The size of the DFS gain, however, depended strongly on $\rho_l$. The maximum $\Delta P$ in the simulation was about 0.07 at $\rho_l$ = 0.70, about 0.21 at $\rho_l$ = 0.90, and about 0.33 at $\rho_l$ = 0.97. This shows that the resistance of DFS to background heterogeneity is not determined by local background subtraction alone. It depends on how well the local background represents the background at the center. When the center and surrounding backgrounds are strongly correlated, their shared background variation can be removed efficiently, so the organism specific signal is easier to preserve even when distractors increase. When $\rho_l$ is low, the surrounding background differs more from the center. Large residual differences then remain after subtraction, and the gain of DFS is smaller.

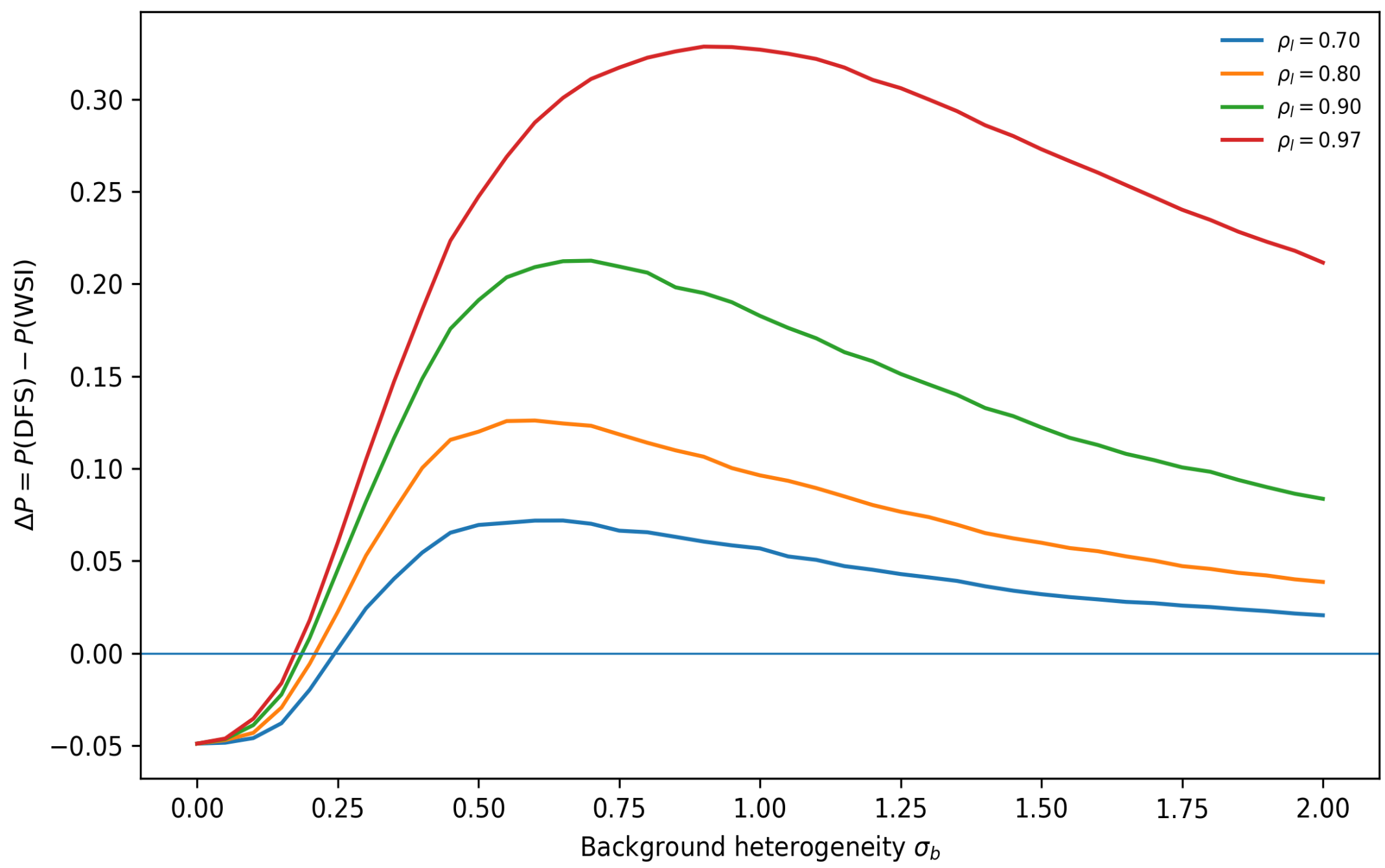

**Figure 8.** Sensitivity to the correlation $\rho_l$ between the central background and the local background, with $B$ fixed at 30. As $\rho_l$ increases, local background subtraction removes background variation more effectively, and the DFS gain becomes larger at moderate or high background heterogeneity.

### 2.8.3 Cost of coarse observation

Finally, we changed the cost $c$ of coarse observation. As $c$ increases, DFS pays a larger fixed cost $M c$ to view all candidate points coarsely. This leaves a smaller budget for fine observation.

For all values $c = 0.02$, 0.05, 0.10, and 0.20, WSI-AI was favored when background heterogeneity was small. The advantage then shifted to DFS as $\sigma_b$ increased. However, as $c$ increased, the crossover moved toward larger values of $\sigma_b$. DFS began to outperform WSI-AI at about $\sigma_b = 0.13$ for $c = 0.02$, 0.19 for $c = 0.05$, 0.23 for $c = 0.10$, and 0.33 for $c = 0.20$. The maximum $\Delta P$ also decreased as $c$ increased. It was about 0.23 for $c = 0.02$ and about 0.15 for $c = 0.20$.

This result agrees with the effect of coarse view cost in Model I. The advantage of DFS does not assume that coarse observation is free. DFS is favored only when the gain in discrimination from local background information exceeds the added cost of the coarse view. When the background is simple, contextual information has little value, so the coarse view cost mainly works against DFS. When background heterogeneity increases and more distractors appear, the information value of selecting candidates with the local background also increases. Within a certain range, this value can offset the cost of coarse observation.

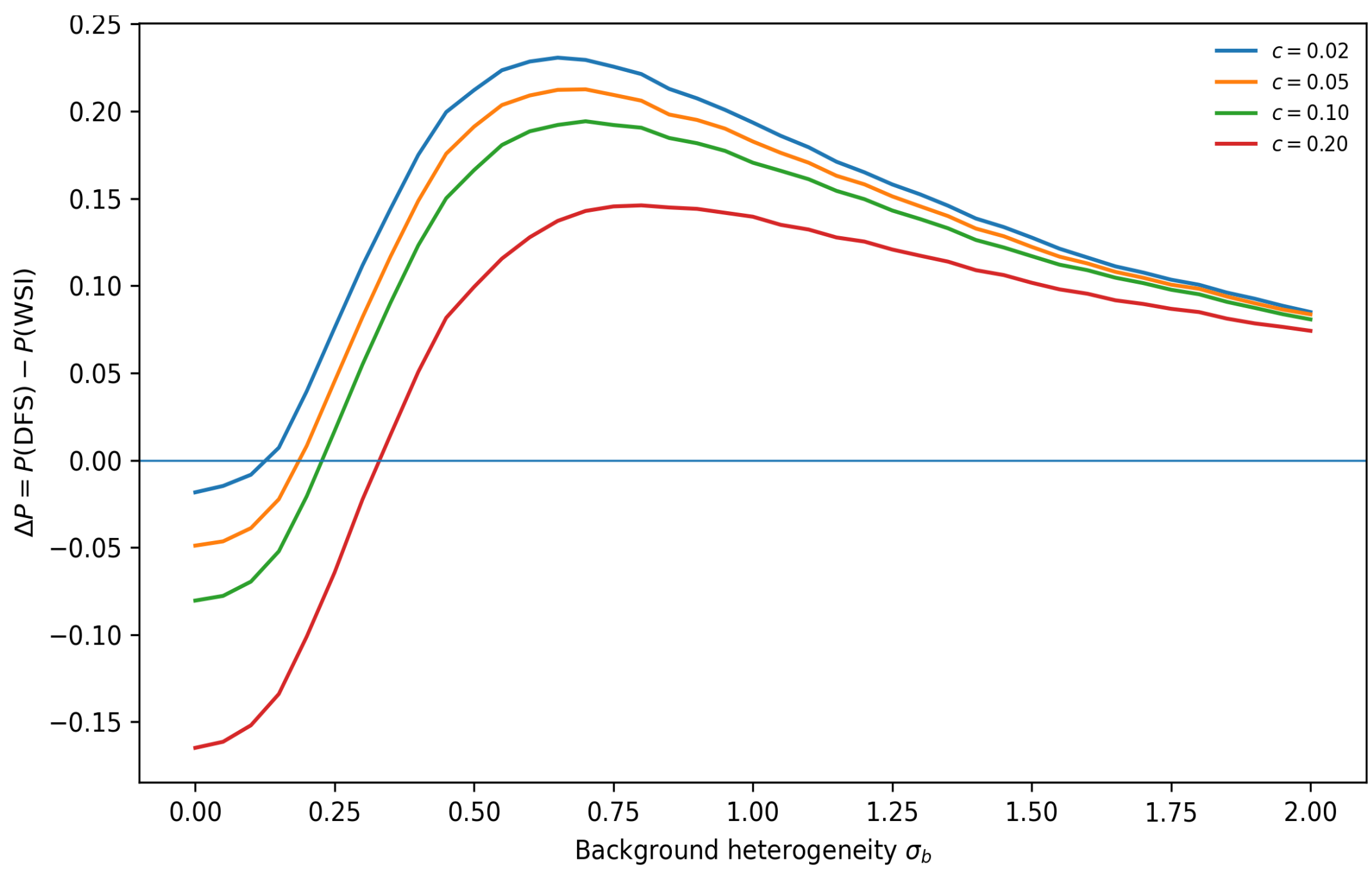

**Figure 9.** Sensitivity to coarse observation cost $c$ with $B$ fixed at 30. As $c$ increases, fewer resources remain for fine observation in DFS. A larger degree of background heterogeneity is therefore needed before DFS gains an advantage, and the maximum gain of DFS also decreases.

### 2.9 Summary of the sensitivity analyses

These sensitivity analyses showed that the main result of Model II was preserved over a wide range of local background correlation $\rho_l$ and coarse view cost $c$. As background heterogeneity increased, the preferred strategy shifted from WSI-AI to DFS. The gain of DFS was generally largest at moderate background heterogeneity and became smaller again when the background became still more complex.

The crossover also depended on the visibility $\kappa$ of the organism signal during coarse observation. When $\kappa$ was large enough for the organism to be recognized already in the coarse view, DFS could retain an advantage even with a uniform background. Therefore, Model II does not mean that DFS must become advantageous whenever background heterogeneity increases. The relative performance of DFS and WSI-AI is instead determined by several factors together. These include how much of the organism signal remains in coarse observation, how well the surrounding background predicts the background at the center, and how much it costs to obtain that information. The region in which DFS had an advantage in Model II can be understood as the range where the information gained by using local context to suppress distractors is larger than the observation cost of the coarse view.

The main sensitivity analyses above compare DFS-AI with the original minimal WSI-AI comparator, which uses absolute redness and does not explicitly apply local background correction. To separate the information value of local contextual normalization from the value of coarse candidate selection, we therefore added a matched control in which WSI-AI uses the same after local background subtraction fine score as DFS-AI while retaining the WSI observation policy and performing no coarse candidate ranking.

### 2.10 WSI control with matched local context

In the context matched control, WSI-AI receives the same fine scale local contextual information as DFS-AI but does not use the coarse view to rank candidates. Under the representative condition $B$ = 30 and $\sigma_b$ = 0.70, the detection probability was 0.1662 for the original WSI-AI comparator, 0.3674 for WSI-AI with matched local context, and 0.3787 for DFS-AI. Thus, of the total DFS advantage of 0.2126 over the original WSI comparator, 0.2012 was accounted for by local contextual normalization and 0.0114 by the additional effect of coarse candidate selection under this condition. Across the parameter grid, the selection only gain was smaller than the contextual gain, although it remained positive over a range of heterogeneous backgrounds (Supplementary Figure S1).

This matched analysis changes the interpretation of Model II in an important way. The large improvement seen in the original comparison between DFS and WSI is primarily evidence for the value of contextual discrimination when distractors arise from background heterogeneity. Coarse candidate selection provides an additional but smaller gain under the baseline parameters. Model II therefore should not be interpreted as showing an intrinsic advantage of observer selection over every context aware WSI method; rather, it separates two sources of information value: local contextual normalization and selective allocation of fine observation resources.

## 3. Model III: Analysis of spatially contiguous lesions under finite resources

### 3.1 Lesion structure and biological interpretation

Model III follows the single rare lesion in Model I and the nonuniform background in Model II. It considers a lesion that itself has a spatially continuous structure. Each point represents one lesion unit. The lesion is a square region made of lesion units with the same lesion state, and every point inside the square has the same color and the same state. The variable in this model is therefore the spatial extent over which similar lesion units continue into neighboring areas.

For each analysis, the side length $s$ of the square is fixed. Across analyses, $s$ is changed among 1, 2, 3, and 4. We interpret $s$ = 1 as a localized lesion and larger $s$ as a more diffuse lesion.

$$n = 7, M = n^2 = 49 \tag{27}$$

$$s \in 1,2,3,4, L = s^2 \in 1,4,9,16 \tag{28}$$

$$Q_s = (n - s + 1)^2 \in 49,36,25,16 \tag{29}$$

Here, $M$ is the total number of lesion units, $L$ is the number of abnormal lesion units, and $Q_s$ is the number of possible positions for a square lesion with side length $s$. Using a 7×7 grid allows direct comparison with $M$ = 49 in Model I.

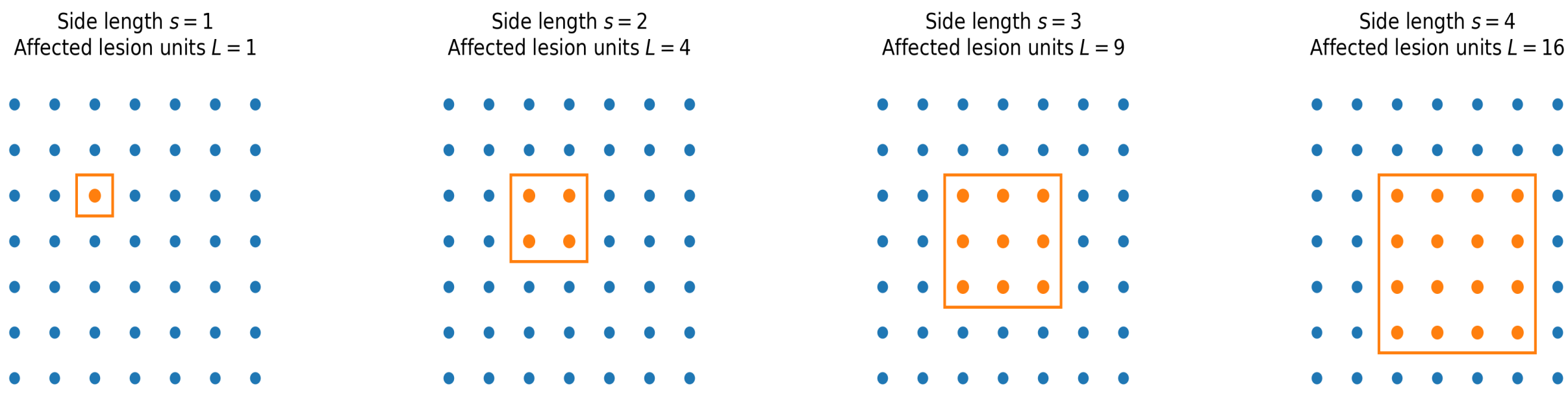

**Figure 10.** Lesion structure in Model III. Each point represents one lesion unit, and only one square lesion is placed. All lesion units inside the square have the same lesion state. Side lengths $s$ = 1, 2, 3, and 4 correspond to 1, 4, 9, and 16 neighboring lesion units being affected. They represent a change from a localized to a more diffuse lesion. Labels in the figure are written in English.

### 3.2 Diagnostic endpoints

Model III separates two endpoints: finding at least one lesion unit and evaluating the entire square lesion. We call the first discovery and the second complete lesion characterization. As a continuous lesion becomes larger, it becomes easier to encounter at least one abnormal unit. However, fine evaluation of the whole lesion requires several lesion units to be handled together. Separating these two endpoints makes the different roles clearer. WSI-AI is strong at search of the entire area, while DFS-AI selects a diagnostic field from a coarse overview.

### 3.3 WSI-AI: Fine observation without location information

Let $B$ be the total observation budget, and let the cost of fine observation of one lesion unit be $c_f = 1$. WSI-AI performs fine observation in an order that does not depend on location and does not use coarse selection in advance. The number of lesion units $k_W$ that can be observed under budget $B$ is

$$k_W = min\left(M, floor\left(B/c_f\right)\right) \quad (30)$$

When the location independent fine observation order is averaged over all coordinates, the probability that WSI-AI observes at least one abnormal lesion unit is given by the following hypergeometric expression.

$$P_{\text{WSI,detect}}(B, s) = 1 - C(M - L, k_{\text{W}})/C(M, k_{\text{W}}) \quad (31)$$

For the whole lesion to count as completely evaluated in this model, all $L = s^2$ abnormal lesion units must be included in fine observation. For this strict endpoint, the probability is

$$P_{\text{WSI,complete}}(B, s) = C(k_{\text{W}}, L)/C(M, L) \quad (32)$$

This definition does not make the general claim that real WSI-AI cannot use local structure. In Model III, WSI-AI is used as a minimal comparator so that we can separate two approaches. One distributes fine resources point by point across the entire area. The other uses a coarse overview to focus fine resources on a continuous field.

### 3.4 DFS-AI: Coarse field scores based on spatial continuity

DFS-AI first performs a low cost coarse observation of all $M$ points. Let $I_i = 1$ for an abnormal lesion unit and $I_i = 0$ for a nonlesion unit. We represent the coarse score $X_i^{(c)}$ at point $i$ with the following minimal model.

$$X_i^{(c)} = \delta I_i + \varepsilon_i, \varepsilon_i \ N(0,1) \quad (33)$$

The parameter $\delta$ is the amount of discriminating information in the coarse view for one lesion unit. The key idea of Model III is that an observer does not have to view each point independently. The observer can use whether a signal continues within a square field of the same scale $s$. We define the structured coarse score $A_q(s)$ of candidate square $q$ as

$$A_q(s) = \left(1/\sqrt{L}\right)\Sigma_{i\in q} X_i^{(c)} \quad (34)$$

We use this score. For the true lesion square, all $L$ lesion units are abnormal, so

$$E[A_{true}(s)] = \delta\sqrt{L} = \delta s, Var[A_{true}(s)] = 1 \quad (35)$$

Thus, even if the coarse discriminability $\delta$ for each lesion unit is unchanged, increasing the side length of the continuous lesion by a factor of $s$ increases the mean separation of the normalized field score to $\delta s$. This is the spatial aggregation gain specific to Model III.

Let the coarse observation cost per point be $c$. The fixed cost of viewing the entire area coarsely is $M c$. Let the cost of fully analyzing one candidate square with fine observation be $L\, c_f$. The number of candidate fields that DFS-AI can analyze in detail, $m_D$, is

$$m_D \;=\; min\left(Q_s, max\left(0, floor\left((B - Mc)/(Lc_f)\right)\right)\right) \tag{36}$$

DFS-AI selects the top $m_D$ candidate squares by $A_q(s)$ and performs fine analysis inside them. Complete lesion characterization is counted as successful if the true square is included among the selected fields. Candidate squares overlap and their scores are correlated. Therefore, $P_{DFS,complete}$ is obtained by Monte Carlo simulation rather than by a closed form that assumes independent ranks.

$$P_{\text{DFS,complete}}(B, s) \;=\; \Pr[\text{rank}(A_{\text{true}}) \;\le\; m_{\text{D}}] \tag{37}$$

$$\Delta P(B, s) \;=\; P_{\text{DFS,complete}}(B, s) \;-\; P_{\text{WSI,complete}}(B, s) \tag{38}$$

**3.5 Numerical simulation**

The representative conditions were $n$ = 7, $M$ = 49, $c_f$ = 1, coarse cost $c$ = 0.05, and coarse discriminability $\delta$ = 0.5 for each lesion unit. $B$ took integer values from 3 to 49. For each condition $s$ = 1, 2, 3, and 4, the true square location was selected uniformly from the $Q_s$ possible positions. Coarse noise for each lesion unit was generated independently from a standard normal distribution. We then calculated $A_q(s)$ for all candidate squares, including overlapping squares. We performed 200,000 Monte Carlo trials for each $s$. The WSI-AI probabilities for discovery and complete characterization were calculated exactly from Equations (31) and (32).

**3.6 Separation of discovery from complete characterization**

The results showed that the gap between finding at least one abnormal unit and evaluating the whole lesion finely became larger as the lesion grew. For example, at $B$ = 30, the WSI-AI discovery probability was 0.612 for $s$ = 1, 0.982 for $s$ = 2, 0.99995 for $s$ = 3, and almost 1 for $s$ = 4. Thus, when a continuous lesion is large, WSI-AI can find the presence of the lesion with almost complete certainty.

At the same $B$ = 30, however, the WSI-AI complete characterization probability, which requires all lesion units to be included in fine observation, was 0.612, 0.129, 0.00696, and 0.000043 for $s$ = 1, 2, 3, and 4, respectively. In the same deterministic 200,000-trial simulations used throughout Model III, the probability that DFS-AI selected the true square field for complete evaluation was 0.733, 0.484, 0.513, and 0.369.

| s | L | WSI: detect ≥1 lesion unit (B = 30) | WSI: complete characterization (B = 30) | DFS: complete characterization (B = 30) |
|---|---|---|---|---|
| 1 | 1 | 0.612 | 0.612 | 0.733 |
| 2 | 4 | 0.982 | 0.129 | 0.484 |
| 3 | 9 | >0.999 | 0.00696 | 0.513 |
| 4 | 16 | ≈1.000 | 0.000043 | 0.369 |

Model III therefore shows that discovery and complete evaluation of lesion extent should not be treated as the same success measure. For a large lesion, entire area fine observation by WSI-AI is very strong for discovery. However, with a limited budget, it is difficult to keep fine information for the entire lesion. If DFS-AI can recognize a continuous pattern in the coarse view, it can focus fine resources on the whole lesion field.

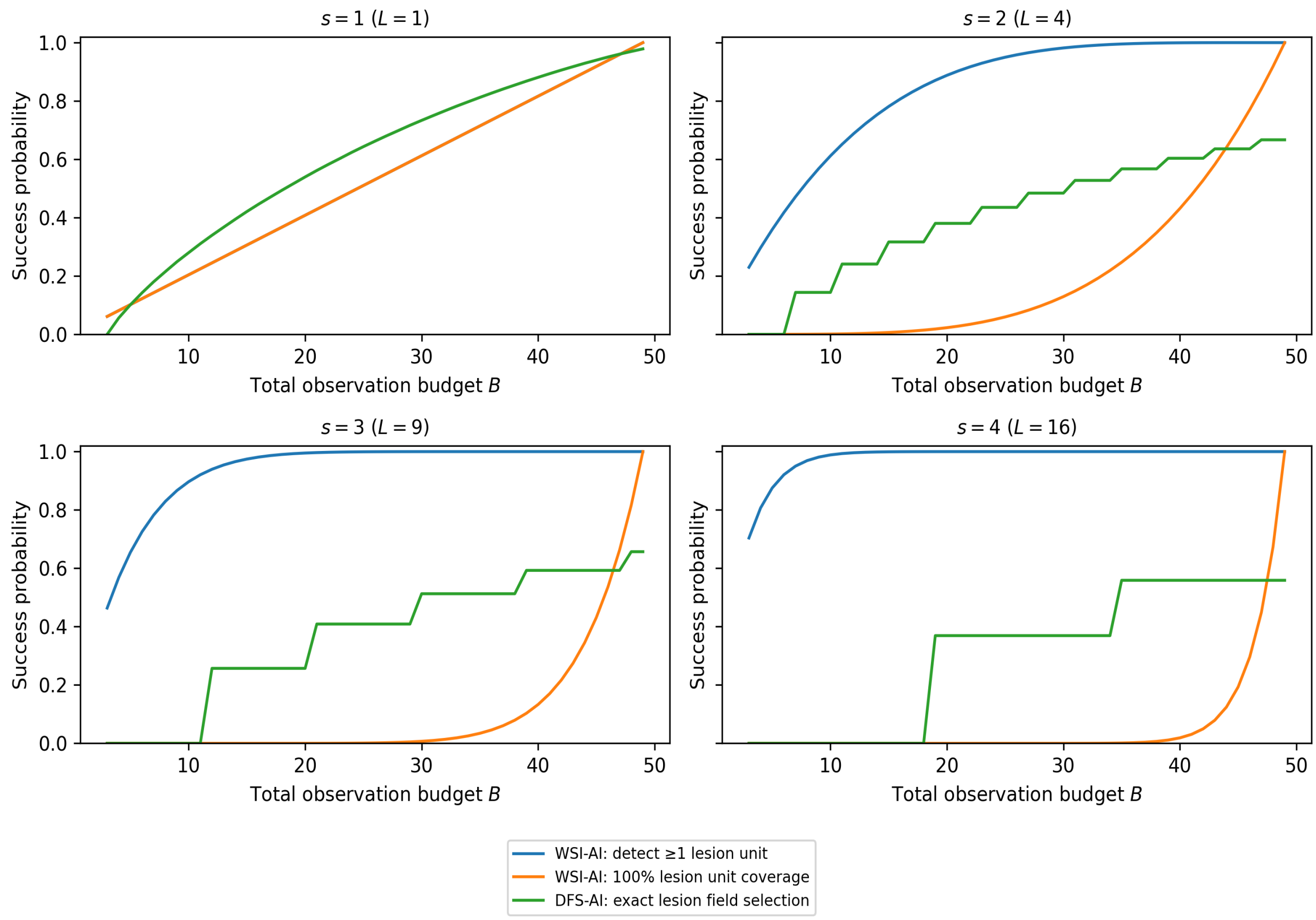

**Figure 11.** Total observation budget $B$ and success probability in Model III. Each panel shows a square lesion with side length $s$ = 1–4. The figure compares three probabilities: WSI-AI discovery of at least one lesion unit, WSI-AI 100% lesion unit coverage, and DFS-AI selection of the exact true square field for complete evaluation. As the lesion becomes larger, WSI-AI discovery rapidly approaches 1. For complete characterization under the primary exact field endpoint, DFS-AI has an advantage over an intermediate resource range.

### 3.7 Resource dependent strategy crossover and lesion size

For the primary exact field complete characterization endpoint, DFS-AI outperformed the pointwise WSI-AI comparator for $B$ = 6–47 when $s$ = 1, $B$ = 7–43 when $s$ = 2, $B$ = 12–46 when $s$ = 3, and $B$ = 19–47 when $s$ = 4. As $s$ increases, the fine cost of one field, $L = s^2$ lesion units, also increases. At very low resources, DFS-AI cannot analyze even one square field in detail, so it has no advantage. For example, when $s$ = 4, the coarse view cost is 2.45 and one 4 × 4 field costs 16 fine observation units, so an integer budget must reach 19 before one field can be analyzed.

In the intermediate resource range, the improvement in candidate ranking from spatial aggregation outweighs the fine analysis cost of a field, and the gain of DFS-AI becomes large. The maximum $\Delta P$ values for $s$ = 1, 2, 3, and 4 were 0.134, 0.401, 0.506, and 0.558, with maxima at approximately $B$ = 24, 27, 30, and 35, respectively. As $B$ approaches $M$ = 49, WSI-AI can observe almost every lesion unit finely. The gain of DFS-AI, which still pays the fixed coarse view cost, then disappears and WSI-AI regains the advantage.

| s | L | B range in which DFS is favored | Maximum ΔP | B at maximum ΔP |
|---|---|---|---|---|

| 1 | 1 | 6–47 | 0.134 | 24 |
|---|---|---|---|---|
| 2 | 4 | 7–43 | 0.401 | 27 |
| 3 | 9 | 12–46 | 0.506 | 30 |
| 4 | 16 | 19–47 | 0.558 | 35 |

Therefore, as in Model I, Model III does not have a single method that is favored throughout the resource range: Low: WSI-AI or not evaluable → Middle: DFS-AI → High: WSI-AI. This gives a strategy crossover as resources change. In Model III, the location of that crossover is further determined by lesion size $s$ and the field cost $L = s^2$.

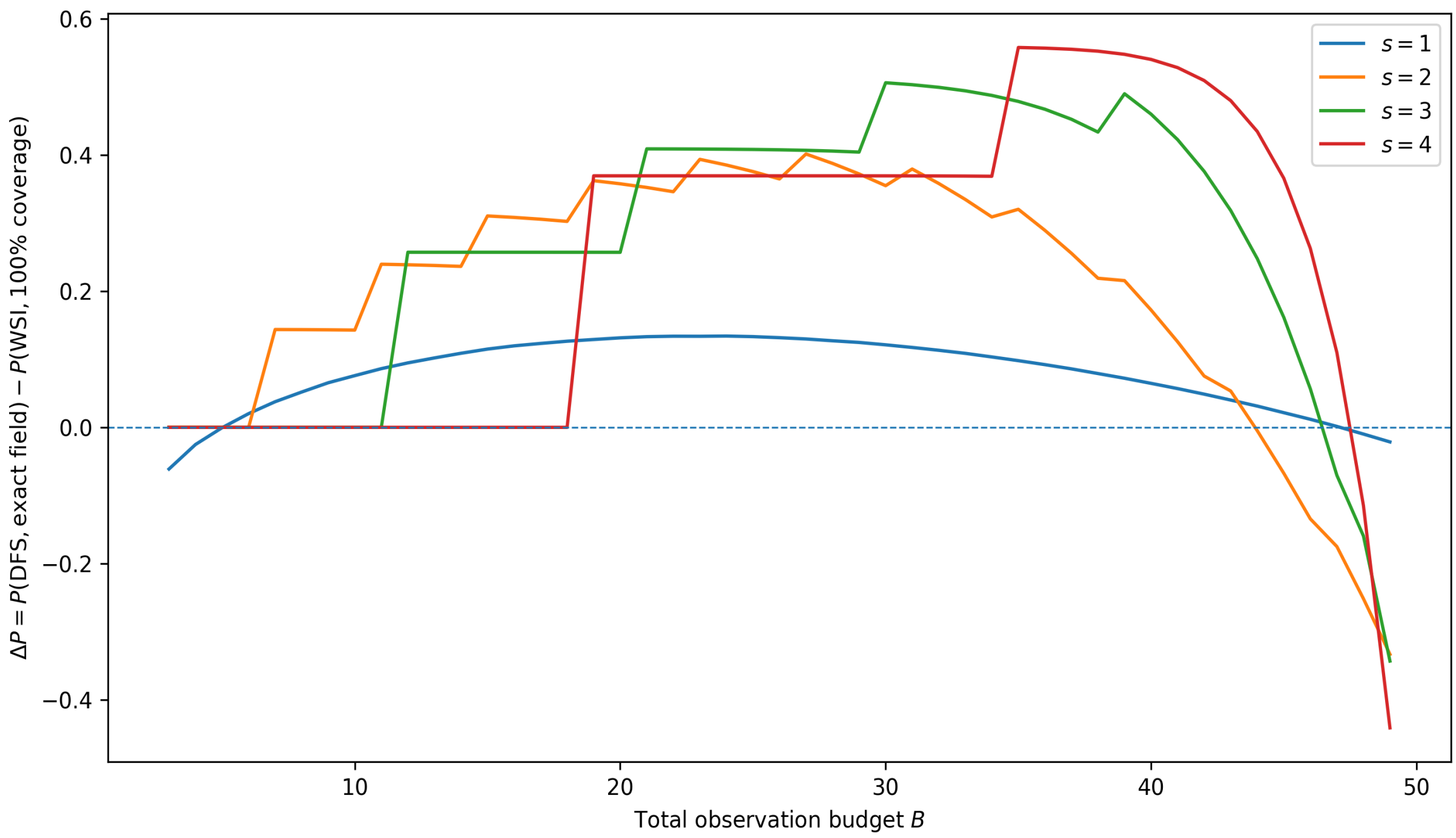

**Figure 12.** DFS advantage in complete characterization in Model III, $\Delta P$ . Values above zero show a DFS-AI advantage. Larger lesions produce a larger maximum gain in the intermediate resource range. The gain disappears at very low resources, where even one field cannot be analyzed, and at high resources, where WSI-AI approaches fine observation of the entire area.

### 3.8 Sensitivity to coarse information and observation cost

With $B$ fixed at 30, we changed the coarse view discriminability $\delta$ per lesion unit from 0 to 2. As $\delta$ increased, $\Delta P$ increased for every $s$, and the increase was stronger for larger $s$. This agrees with the equation $E[A_{true}] = \delta s$. As the number of continuous lesion units increases, the same weak signal can be added together at the field level.

When coarse cost $c$ increased from 0 to 0.30, the number of candidate fields $m_D$ that DFS-AI could analyze finely decreased in steps, and $\Delta P$ decreased. For $s = 1$, the DFS advantage disappeared at about $c = 0.20$. For $s > 1$, the field cost is discrete, so the change was stepwise. The gain disappeared when the coarse view cost became so large that even one candidate field could no longer receive fine analysis.

Interestingly, for $s > 1$, a small DFS gain can remain even when $\delta = 0$. This occurs because allocating fine resources in coherent square fields can increase the probability of covering a spatially continuous lesion compared with point-by-point allocation, even when the coarse score itself is uninformative. The Model III gain can therefore be decomposed into structured field allocation and additional ranking from coarse structural information.

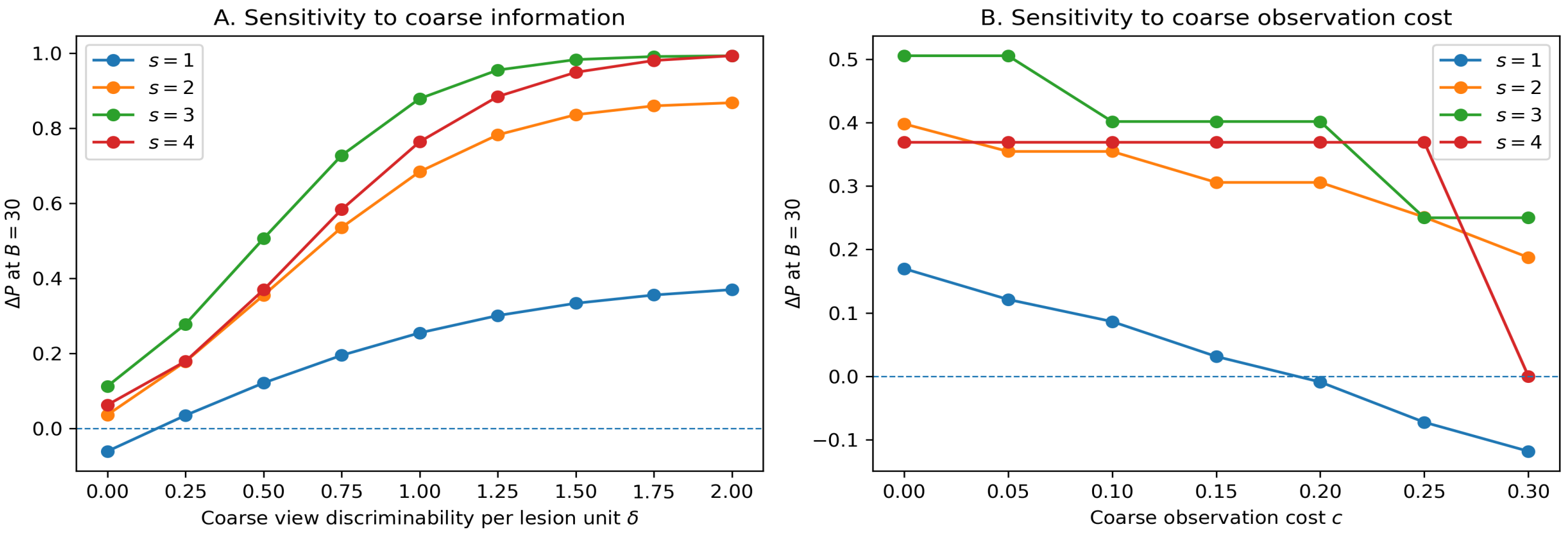

**Figure 13.** Sensitivity analysis in Model III. A: $\Delta P$ as a function of coarse view discriminability $\delta$ at $B = 30$. As $\delta$ increases, the gain of DFS-AI increases. The effect of spatial aggregation is stronger for larger lesions. *B*: $\Delta P$ as a function of coarse observation cost *c*at *BB* = 30 and $\delta = 0.5$. As *c* increases, the number of fields that can receive fine analysis decreases, and the DFS gain becomes smaller.

### 3.9 Control matched for field structure and lesion unit coverage endpoints

To determine whether the DFS advantage in Model III was caused simply by comparing DFS allocation by field with pointwise WSI allocation, we added a field comparator matched for structure. This control performs no coarse observation and spends the entire budget on randomly selected $s \times s$ fields. For complete characterization, its success probability is therefore the fraction of candidate square fields that can be analyzed within the budget. This control isolates the benefit of allocation by field itself from the additional benefit of ranking fields by the coarse structural score.

At $B = 30$, the complete characterization probabilities for pointwise WSI-AI, the random field control matched for structure, and DFS-AI were respectively 0.612, 0.612, and 0.733 for $s = 1$; 0.129, 0.194, and 0.484 for $s = 2$; 0.00696, 0.120, and 0.513 for $s = 3$; and 0.000043, 0.0625, and 0.369 for $s = 4$. Thus, allocation by field alone explained part of the gain for larger lesions, but a substantial advantage remained after matching the unit of fine analysis. For example, at $s = 3$ the gain from allocation by field was 0.113, whereas the additional gain from coarse structural selection was 0.393 (Supplementary Figure S2A-*B*).

We also tested endpoint symmetry at the lesion unit level. In addition to the primary exact field endpoint, 100% coverage was counted as successful whenever the union of all fields selected by DFS covered every lesion unit, regardless of whether the true square itself was selected exactly. At $B = 30$, the symmetric 100% coverage probability for WSI-AI versus DFS-AI was 0.129 versus 0.495 for $s = 2$, 0.00696 versus 0.5266 for $s = 3$, and 0.000043 versus 0.369 for $s = 4$. The endpoint was then relaxed to at least 90% or at least 80% lesion unit coverage. At $s = 3$, the corresponding WSI-AI versus DFS-AI probabilities were 0.0070 versus 0.5266 for at least 90% coverage and 0.0611 versus 0.5725 for at least 80% coverage. At $s = 4$, they were 0.00092 versus 0.369 and 0.0430 versus 0.369, respectively. The Model III conclusion therefore persists under symmetric 100% coverage and relaxed 90% and 80% lesion unit endpoints (Supplementary Figure S2C-E).

### 3.10 Relationship to Models I and II

In Model I, the value of DFS-AI came from spatial localization information obtained through coarse observation. In Model II, contextual discrimination based on local background information was added in the presence of background heterogeneity and distractors. Model III adds another source of information: several neighboring lesion units can form one continuous lesion structure. The information available to DFS-AI can therefore be organized as

spatial localization + contextual discrimination + structural aggregation

These are three layers of information. In Model III, the amplification to $\delta s$ means that even a weak finding in each lesion unit can be easier to recognize as a diagnostic field in the coarse view when the finding continues at the same spatial scale. This corresponds to tasks such as vasculitis, the border of necrosis, and an invasive front, where the distribution pattern itself can carry more diagnostic information than a single local cell.

At the same time, Figure 11 shows that if the goal is only to find at least one part of the lesion, WSI-AI has very high sensitivity as the lesion becomes larger. Model III therefore supports a division of roles: WSI-AI for lesion discovery, and DFS-AI for evaluating the extent and relationships of a structure after it has been found.

### 3.11 Conclusions and limitations of Model III

The main result of Model III is that spatial continuity can increase the amount of information at the coarse field level even when the local finding itself is unchanged. This raises the value of Diagnostic Field Selection under limited resources. Because the mean normalized true field score increases as $\delta\sqrt{L} = \delta s$, lesion size is not only the amount of disease; it also acts as structural information that helps field selection. The control matched for field structure shows that part of the advantage comes from allocating fine resources in coherent fields, but a substantial additional gain remains when coarse structural information is used to rank those fields. The same conclusion is preserved when success is defined symmetrically by 100% lesion unit coverage and when the threshold is relaxed to 90% or 80% coverage.

This is a minimal model that assumes a square grid, one lesion, a known analysis scale $s$, and uniform lesion strength. Real vasculitis lesions can have irregular shapes, uneven vessel density, several lesions, and variation in lesion strength. If WSI-AI used similar multiscale spatial aggregation or local structure recognition, the difference from DFS-AI shown here could become smaller. These results therefore do not assign an unavoidable advantage to one image acquisition method. They isolate the information value of an observer recognizing a continuous structure in a coarse view and selecting a field for fine analysis under limited resources.

In summary, Model III shows the following complementary relationship:

WSI-AI: global discovery / DFS-AI: structure guided complete characterization

WSI-AI is favored when the main endpoint is to find lesions without missing them. DFS-AI may be favored when the main endpoint is to evaluate the spatial extent and continuity of a lesion with limited fine resources.

## 4. Conditions that favor each strategy

### 4.1 Conditions that favor DFS-AI

DFS-AI is more likely to be useful when several of the following conditions are present: (i)Diagnostic information is not distributed evenly within the WSI; (ii) Observers can identify the fields needed for diagnosis with reasonable consistency; (iii) Relationships across several fields and magnifications are important for diagnosis; (iv) There is large variation with low diagnostic relevance, such as normal tissue, staining differences, tissue amount, and artifacts: (v) The number of cases available to train a fully automatic selector is limited.

For histologic type classification, tissue patterns such as glands, nests, cords, and papillary structures must be combined with high magnification cell features. Invasion cannot be judged only by the presence of tumor cells. The relationship between the lesion and surrounding tissue, including contact, destruction, and stromal response, must also be evaluated. Lesions also need to be understood through the relationship among cell type, tissue compartment, distribution, tissue injury, and time phase, not just by cell count. In such tasks, observers can choose fields and magnifications based on a diagnostic hypothesis and use AI for local classification, measurement, and comparison. This can focus information resources on the most relevant parts. As shown in Model III, when similar lesions extend continuously into neighboring lesion units, recognizing the structure in a coarse view and selecting a field can itself help focus fine analysis resources.

In the main diagnostic benchmark for PathChat, pathologists selected salient ROIs from the WSI and entered them into the model. This can be viewed as a real example of DFS style use of AI [12]. This does not mean that PathChat itself should be called DFS. It simply shows that observer selection of the fields given to AI is a practical way to use the model.

### 4.2 Conditions that favor WSI-AI

WSI-AI is strong when finding the target location itself is the main task, especially when an observer cannot limit the location in advance. This can occur in a large surgical section when it is not even known whether a lesion is present. For a target positive case, let $p$ be the probability that the target is included in the AI analysis, and let $s$ be the conditional sensitivity after the target has been included. A simplified overall sensitivity is then

$$\text{Sensitivity} = p\,s \tag{39}$$

With observer field selection in advance, a target outside the selected fields cannot be detected no matter how accurate the later AI analysis is. Coverage $p$ therefore sets the upper limit of overall sensitivity. If WSI-AI can search all tissue at enough resolution and make $p$ close to 1, it has a major advantage for finding rare targets.

The presence or absence of lymph node metastasis, especially a small metastatic focus, is a typical example. Here, the main information processing task is to find a target whose location is unknown before interpreting what the candidate is. Studies such as CAMELYON16 have shown that WSI-AI can achieve high performance in this type of search of the entire area [11].

## 5. Discussion

The central contribution of this study is not the claim that DFS-AI is intrinsically superior to WSI-AI, but the identification of a principle by which the preferred strategy changes with available resources. Broad coverage is favored when selection information is weak, when coarse observation is costly, or when the main endpoint is discovery of an unknown target. Selective analysis becomes favorable when an overview acquired at relatively low cost provides enough localization, contextual, or structural information to concentrate costly fine observation. When resources become sufficient for broad observation at fine resolution, the value of selection diminishes again. Models I–III identify three mechanisms that shift this crossover boundary: localization information, contextual discrimination against distractors, and structural aggregation across neighboring lesion units.

From this view, DFS is best understood not as a method that uses a small ROI, but as a method in which an observer distributes the information rate across space and magnification according to the diagnostic task [14, 15]. This is especially important for relational tasks such as histologic type, invasion, and inflammation. The key is not to compress everything into one field, but to keep the relationships among the several fields that are needed.

In contrast, a task such as detecting a small lymph node metastasis in a surgical specimen is a clear counterexample to DFS. If the cost of missing a target is high, observer selection in advance may reduce coverage. In that case, it can be worth paying the computing cost of processing regions with low relevance in order to search the entire area. Information selection for interpretation and discovery of a target with an unknown location should therefore be discussed separately. In this situation, a suitable workflow may be WSI analysis supported by prior learning from the primary lesion, followed by observer review of the result.

The quantities $c_N$ and $d_D(C)$ used in the restricted model are theoretical quantities, not directly observed physical measurements. They may be estimated in an operational way from factors such as the background dependence of WSI-AI, experiments that remove or exchange background regions, and the difference in log loss between DFS and the full WSI. What counts as low relevance information $N$ also depends on the diagnostic task. A region with low relevance for histologic type diagnosis may still be important for prognosis or for measuring the total amount of disease.

The matched controls refine this interpretation. In Model II, providing WSI-AI with the same local background correction as DFS-AI removes most of the original performance gap, showing that the dominant information gain in that model is contextual discrimination rather than selection alone. Coarse candidate ranking still adds a smaller gain under heterogeneous backgrounds. In Model III, by contrast, the DFS advantage remains substantial after the fine analysis unit is matched by allowing the comparator to sample whole square fields. The remaining difference therefore reflects the value of using coarse structural information to rank coherent fields, not merely the fact that DFS analyzes contiguous regions. The result also persists when 100% success is defined symmetrically by lesion unit coverage rather than exact square selection, and when the threshold is relaxed to 90% or 80% coverage.

These results also clarify the scope of the proposed framework. Previous work on pathologist viewing behavior, learned viewing policies, and selective WSI processing from coarse to fine resolution motivates the individual mechanisms considered here [10, 13, 16]. The present study instead asks which strategy should be preferred under different information and resource conditions: broad whole slide coverage or selective analysis of Diagnostic Fields. The matched controls are important because they separate the value of contextual or structural information from the additional value of selection itself. This reduces the possibility that the conclusions about resource allocation arise only from an asymmetric comparator.

Research such as Pathology-CoT [13], which trains AI on the observers' own viewing behavior, includes a different problem from the comparison in this study. If AI learns past observer viewing strategies, changes future observer viewing through its own suggestions, and those changed behaviors are then returned to the training data, the process becomes a recursive system in which later training data depend on an earlier AI. We do not include this recursive learning in the static comparison of WSI-AI and DFS-AI. In our comparison, observer viewing strategy remains outside the AI, and AI is treated as an image analysis tool.

The numerical analyses were generated with deterministic Python scripts. Model II used 20,000 paired Monte Carlo trials for each parameter condition, with common random numbers across the original WSI-AI comparator, WSI-AI with matched local context, and DFS-AI. Model III used 200,000 Monte Carlo trials for each lesion side length, with deterministic seeds derived from the base seed 20260908. Exact WSI probabilities were calculated analytically from the corresponding combinatorial or hypergeometric expressions. The same Model III simulation output was used for the main text, tables, Figures 11–13, and Supplementary Figure S2 to avoid numerical differences between runs.

In pathology AI under limited resources, giving the whole WSI as input does not by itself guarantee the best use of diagnostic information. WSI-AI provides coverage of the entire area, but it uses limited resources to process, compress, and combine regions with low relevance. DFS-AI can focus resources on diagnostically relevant fields, but it carries the risk that an observer may miss an important field during selection. The relative advantage of the two methods therefore depends on which is larger: information loss in WSI-AI from background processing and aggregation, or information loss in DFS from selection.

DFS-AI is more likely to be useful when diagnostic information is unevenly distributed, spans several scales, depends on relationships among findings, and requires observer knowledge. In such tasks, use of DFS-AI by a pathologist may improve the practical value of AI assistance. By contrast, WSI-AI is more likely to be useful when the main task is to find a rare target without missing it, as in the detection of a small lymph node metastasis. In such settings, use of WSI-AI by a pathologist may be advantageous. In clinical practice, the form of AI assistance should therefore be selected by the pathologist according to the information structure of the diagnostic task.

## 6. Declarations

Ethics approval: Not applicable. This study used mathematical models and synthetic simulations only and involved no human participants, patient data, human specimens, or animal experiments.

Data availability: No patient data were analyzed. Numerical results were generated from the mathematical models and synthetic simulations described in this manuscript.

Code availability: The Python script Diagnostic_Field_Selection_simulation_code.py, used to generate the numerical simulations, matched controls, and revised figures, is provided as supplementary material.

Author contributions: T.T. conceived the study, developed the models, performed the analyses, interpreted the results, prepared the figures, and wrote the manuscript.

Funding: This study received support from an AMED project on familial carcinogenesis.

Competing interests: None.

## Supplementary material

### Supplementary Figure S1. Control with matched local context in Model II

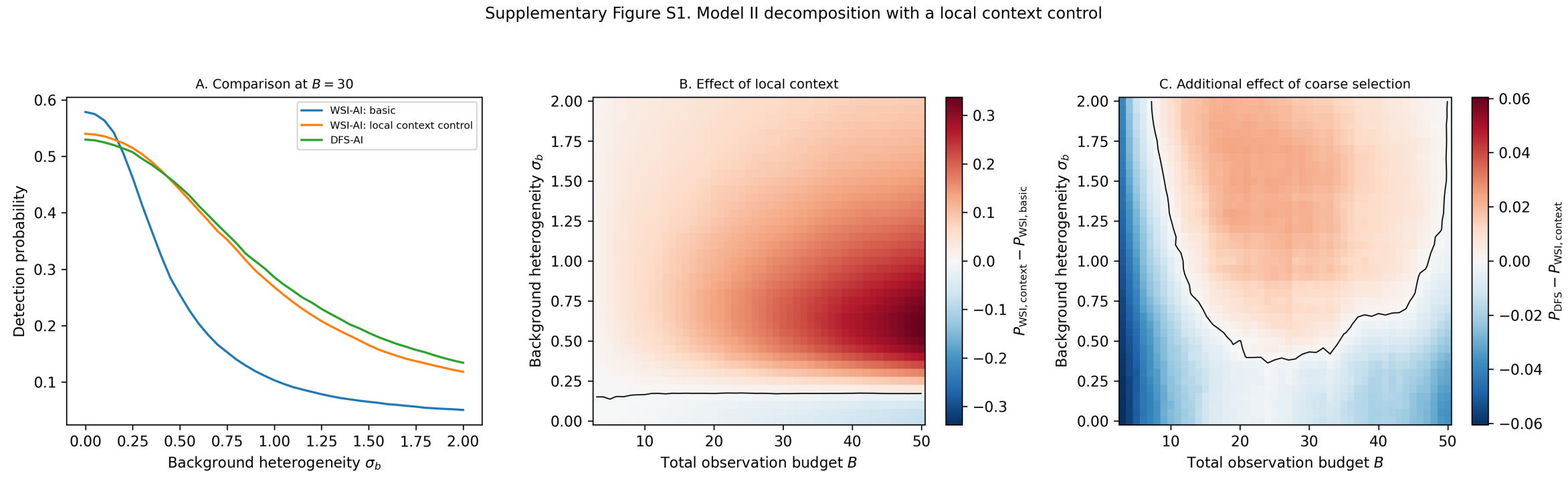

Supplementary Figure S1. Control with matched local context in Model II. A: Detection probability at B = 30 for the original WSI-AI comparator, a WSI-AI control that uses the same local background subtraction as DFS-

AI but does not use coarse candidate ranking, and DFS-AI. B: Effect of local context, defined as P(WSI-context) − P(WSI-basic), across total observation budget B and background heterogeneity σb. C: Additional effect of coarse selection after matching local context, defined as P(DFS) − P(WSI-context). Model II used 20,000 paired Monte Carlo trials for each parameter condition, with common random numbers across the three methods. This control separates the value of local contextual normalization from the additional value of coarse candidate selection.

**Supplementary Figure S2. Control matched for field structure and lesion unit coverage sensitivity in Model III**

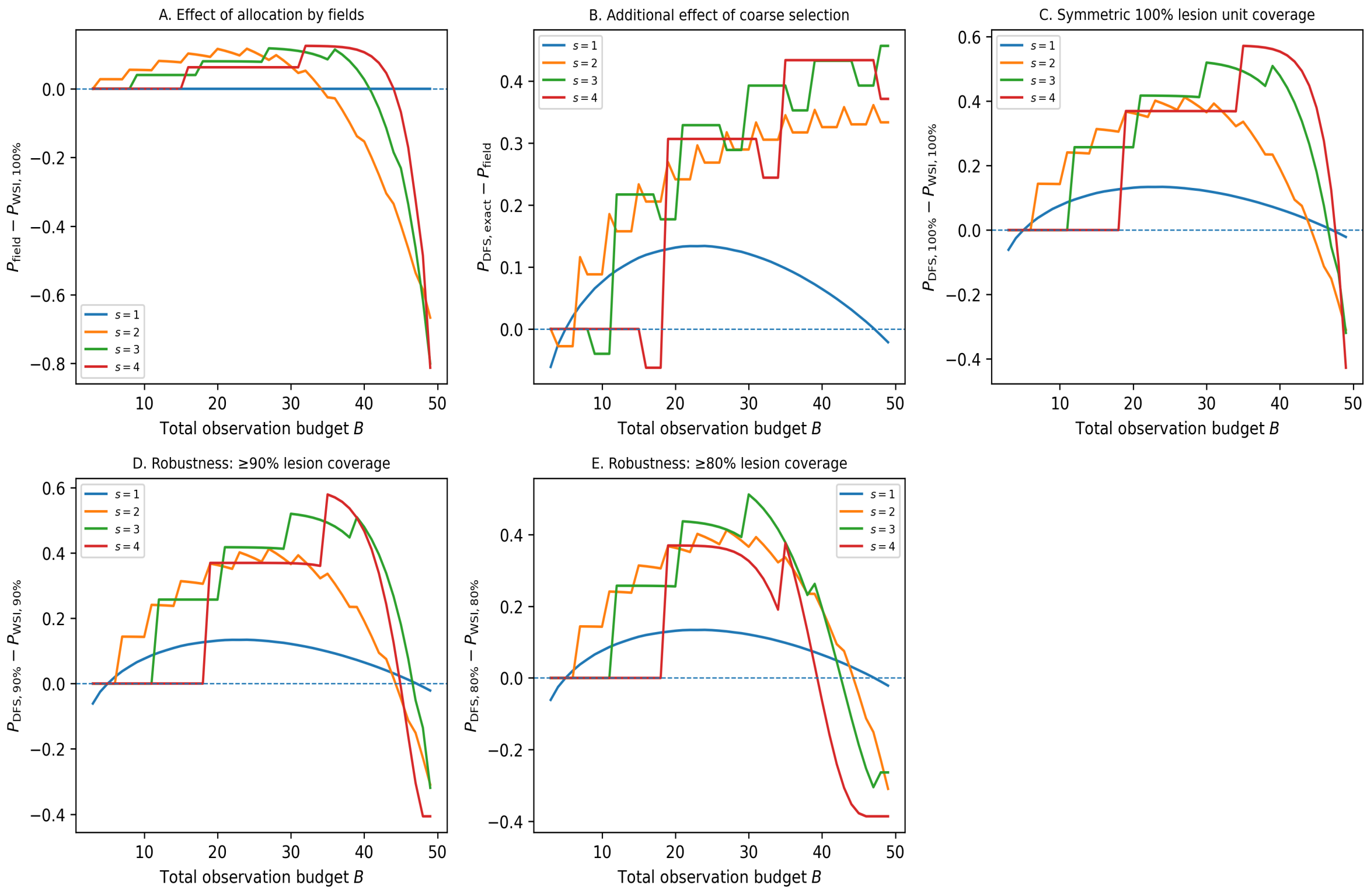

Supplementary Figure S2. Field control and lesion unit coverage sensitivity in Model III. A: Effect of allocating fine analysis by fields alone, defined as P(random field sampling with matched structure) − P(pointwise WSI 100% coverage). The control performs no coarse observation and uses the entire budget for randomly selected s × s fields. B: Additional effect of coarse selection after matching the unit of fine analysis, defined as P(DFS exact field selection) − P(random field selection with matched structure). C: Symmetric 100% lesion unit coverage. DFS is counted as successful when the union of selected fields covers all lesion units, irrespective of whether the exact square is selected. D–E: Advantage of DFS-AI over pointwise WSI-AI when successful characterization is relaxed to at least 90% or at least 80% lesion unit coverage. DFS probabilities were estimated from 200,000 Monte Carlo trials for each lesion side length s. WSI probabilities and the random field control were calculated analytically when applicable.

```
"""
Reproducible figure code for:
Resource-Constrained Diagnostic Field Selection

Generates main Figures 1-13 and Supplementary Figures S1-S2 in manuscript order.

Default settings reproduce the final analysis:
  Model II: 20,000 paired Monte Carlo trials per parameter condition
  Model III: 200,000 Monte Carlo trials per lesion side length
  Base random seed: 20260908

For a fast syntax/plot test only, run:
  QUICK=1 python Diagnostic_Field_Selection_all_figure_code.py

Outputs are written to ./figure_output/ as PNG and PDF.

IMPORTANT MODEL-II IMPLEMENTATION NOTE
-------------------------------------
The numerical code used for the final Model II figures uses:
    sigma_c       = 0.70   # central coarse-observation noise
    sigma_local_c = 0.40   # local-background coarse-observation noise
    sigma_f       = 0.20
    sigma_local_f = 0.15
These are the values used in the calculations below. They should be kept
consistent with the Methods text when the manuscript is finalized.
"""

from pathlib import Path
import math
import os
import numpy as np
import pandas as pd
import matplotlib.pyplot as plt
from matplotlib.colors import TwoSlopeNorm
from matplotlib.patches import Rectangle, Circle
from scipy.special import roots_hermitenorm
```

```
from scipy.stats import norm, binom, hypergeom

#
===============================================================
===
# Global settings
#
===============================================================
===
OUT = Path("figure_output")
OUT.mkdir(exist_ok=True)
SEED = 20260908
QUICK = os.getenv("QUICK", "0") == "1"

N2 = 1_000 if QUICK else 20_000
N3 = 5_000 if QUICK else 200_000

plt.rcParams.update({
    "font.size": 10,
    "axes.titlesize": 11,
    "axes.labelsize": 10,
    "legend.fontsize": 8,
    "figure.dpi": 120,
    "savefig.dpi": 600,
    "axes.spines.top": True,
    "axes.spines.right": True,
})


def save_figure(fig, stem):
    fig.savefig(OUT / f"{stem}.png", dpi=600, bbox_inches="tight")
    fig.savefig(OUT / f"{stem}.pdf", bbox_inches="tight")
    plt.close(fig)



#
```

```
============================================================
===
# Figure 1. Model I observation strategies
#
============================================================
===
def figure1_model1_schematic():
    n = 7
    lesion = (2, 4)    # row, column; schematic location only
    x = np.arange(n)
    y = np.arange(n)[::-1]

    fig, axes = plt.subplots(1, 3, figsize=(14.5, 5.0))
    titles = [
        "A     Model I ground truth",
        "B     WSI: fine trace",
        "C     DFS: coarse view → focus",
    ]

    for ax, title in zip(axes, titles):
        ax.set_title(title, loc="left", fontweight="bold")
        ax.set_xlim(-0.8, 6.8)
        ax.set_ylim(-0.8, 6.8)
        ax.set_aspect("equal")
        ax.set_xticks([])
        ax.set_yticks([])
        for spine in ax.spines.values():
            spine.set_linewidth(1.2)

    # Panel A
    for r in range(n):
        for c in range(n):
            color = "tab:red" if (r, c) == lesion else "tab:blue"
            axes[0].scatter(c, n - 1 - r, s=55, c=color, zorder=3)

    # Panel B: serpentine full trace, with 15 fine-inspected sites highlighted
```

```
    trace = []
    for r in range(n - 1, -1, -1):    # starts at lower left
        cols = range(n) if (n - 1 - r) % 2 == 0 else range(n - 1, -1, -1)
        for c in cols:
            trace.append((r, c))
    trace_xy = [(c, n - 1 - r) for r, c in trace]
    axes[1].plot([p[0] for p in trace_xy], [p[1] for p in trace_xy], lw=1.1, alpha=0.75)
    fine_sites = set(trace[:15])
    for r in range(n):
        for c in range(n):
            color = "tab:red" if (r, c) == lesion else "tab:blue"
            axes[1].scatter(c, n - 1 - r, s=55, c=color, zorder=3)
            if (r, c) in fine_sites:
                axes[1].add_patch(Circle((c, n - 1 - r), 0.18, fill=False, ec="0.30", lw=1.3,
zorder=4))

    # Panel C: all sites receive a light coarse-view halo; 12 selected sites get a dark halo.
    selected = {
        (0, 3), (0, 5), (1, 2), (1, 5), (2, 1), (2, 4),
        (3, 4), (3, 5), (4, 4), (5, 4), (6, 1), (6, 2),
    }
    for r in range(n):
        for c in range(n):
            yy = n - 1 - r
            axes[2].add_patch(Circle((c, yy), 0.15, fill=False, ec="0.75", lw=0.9,
zorder=1))
            if (r, c) in selected:
                axes[2].add_patch(Circle((c, yy), 0.20, fill=False, ec="0.25", lw=1.4,
zorder=2))
            color = "tab:red" if (r, c) == lesion else "tab:blue"
            axes[2].scatter(c, yy, s=55, c=color, zorder=3)

    fig.text(
        0.5, 0.035,
        "Red = lesion; blue = normal. In the calculation, the lesion is moved through all 49
coordinates.¥n"
```

```
        r"Representative budget $B=15$: WSI fine-inspects 15 sites; DFS first surveys all 49
sites coarsely "
        r"(cost 2.45) and then fine-inspects 12 ranked sites.",
        ha="center", va="bottom", fontsize=9,
    )
    fig.tight_layout(rect=[0, 0.10, 1, 1])
    save_figure(fig, "Figure_01_Model_I_schematic")


#
=============================================================
===
# Model I mathematics used by Figures 2-4
#
=============================================================
===
M1 = 49
C_FINE_1 = 1.0
C_COARSE_1 = 0.05


def p_wsi_model1(B, M=M1, c_f=C_FINE_1):
    k_w = min(M, max(0, int(math.floor(B / c_f))))
    return k_w / M


def m_d_model1(B, c=C_COARSE_1, M=M1, c_f=C_FINE_1):
    return min(M, max(0, int(math.floor((B - M * c) / c_f))))


# High-order Gauss-Hermite quadrature for the exact normal expectation in Model I.
# 240 nodes reproduces the direct integral to approximately 1e-9 in the representative case.
_GH_X, _GH_W = roots_hermitenorm(240 if not QUICK else 80)

def p_dfs_model1(B, delta, c=C_COARSE_1, M=M1, c_f=C_FINE_1):
    """Coordinate-averaged DFS detection probability evaluated by Gaussian quadrature."""
```

```
    m_d = m_d_model1(B, c=c, M=M, c_f=c_f)
    if m_d <= 0:
        return 0.0
    if m_d >= M:
        return 1.0

    # Z_1 = delta + X, X ~ N(0,1). Conditional on Z_1=z, the number of
    # normal scores exceeding z is Bin(M-1, 1-Phi(z)).
    z = delta + _GH_X
    q = 1.0 - norm.cdf(z)
    conditional = binom.cdf(m_d - 1, M - 1, q)
    return float(np.sum(_GH_W * conditional) / np.sqrt(2.0 * np.pi))


#
================================================================
===
# Figure 2. Model I budget-response curve
#
================================================================
===
def figure2_model1_budget():
    B = np.arange(3, 50)
    delta = 0.5
    c = 0.05
    p_w = np.array([p_wsi_model1(b) for b in B])
    p_d = np.array([p_dfs_model1(b, delta, c) for b in B])

    fig, ax = plt.subplots(figsize=(6.5, 4.8))
    ax.plot(B, p_w, label="WSI-AI: fine trace")
    ax.plot(B, p_d, label="DFS-AI: coarse overview → ranked fine analysis")
    ax.set_title("Model I: performance averaged over lesion coordinates")
    ax.set_xlabel(r"Total observation budget $B$")
    ax.set_ylabel("Probability of detecting the lesion")
    ax.set_xlim(3, 49)
    ax.set_ylim(0, 1.02)
```

```
    ax.legend(frameon=True)
    fig.tight_layout()
    save_figure(fig, "Figure_02_Model_I_budget_response")


#
=============================================================
===
# Figure 3. Model I phase/crossover map over B and delta
#
=============================================================
===
def figure3_model1_delta_map():
    B = np.arange(3, 50)
    delta_grid = np.linspace(0.0, 2.0, 81 if not QUICK else 21)
    Z = np.empty((len(delta_grid), len(B)))
    p_w = np.array([p_wsi_model1(b) for b in B])

    for i, d in enumerate(delta_grid):
        Z[i, :] = np.array([p_dfs_model1(b, d, C_COARSE_1) for b in B]) - p_w

    vmax = max(abs(np.nanmin(Z)), abs(np.nanmax(Z)))
    fig, ax = plt.subplots(figsize=(6.3, 5.2))
    mesh = ax.pcolormesh(
        B, delta_grid, Z, shading="auto", cmap="RdBu_r",
        norm=TwoSlopeNorm(vmin=-vmax, vcenter=0, vmax=vmax),
    )
    ax.contour(B, delta_grid, Z, levels=[0], colors="0.2", linewidths=1.0)
    ax.set_title("Model I: DFS advantage over WSI")
    ax.set_xlabel(r"Total observation budget $B$")
    ax.set_ylabel(r"Coarse view discriminability $¥delta$")
    cb = fig.colorbar(mesh, ax=ax)
    cb.set_label(r"$¥Delta P=P(¥mathrm{DFS})-P(¥mathrm{WSI})$")
    ax.text(0.62, 0.90, r"Red: DFS advantage ($¥Delta P>0$)", transform=ax.transAxes,
            bbox=dict(facecolor="white", alpha=0.8, edgecolor="0.7"), fontsize=8)
    ax.text(0.62, 0.81, r"Blue: WSI advantage ($¥Delta P<0$)", transform=ax.transAxes,
```

```
            bbox=dict(facecolor="white", alpha=0.8, edgecolor="0.7"), fontsize=8)
    fig.tight_layout()
    save_figure(fig, "Figure_03_Model_I_delta_crossover_map")


#
============================================================
===
# Figure 4. Model I sensitivity to coarse-view cost
#
============================================================
===
def figure4_model1_cost_sensitivity():
    B = np.arange(3, 50)
    costs = np.linspace(0.0, 0.20, 81 if not QUICK else 21)
    delta = 0.5
    Z = np.empty((len(costs), len(B)))
    p_w = np.array([p_wsi_model1(b) for b in B])
    for i, c in enumerate(costs):
        Z[i, :] = np.array([p_dfs_model1(b, delta, c) for b in B]) - p_w

    fig, axes = plt.subplots(1, 2, figsize=(11.0, 4.6))
    vmax = max(abs(np.nanmin(Z)), abs(np.nanmax(Z)))
    mesh = axes[0].pcolormesh(
        B, costs, Z, shading="auto", cmap="RdBu_r",
        norm=TwoSlopeNorm(vmin=-vmax, vcenter=0, vmax=vmax),
    )
    axes[0].contour(B, costs, Z, levels=[0], colors="0.2", linewidths=1.0)
    axes[0].set_title("A. Budget × coarse observation cost")
    axes[0].set_xlabel(r"Total observation budget $B$")
    axes[0].set_ylabel(r"Coarse observation cost per site $c$")
    cb = fig.colorbar(mesh, ax=axes[0])
    cb.set_label(r"$¥Delta P=P(¥mathrm{DFS})-P(¥mathrm{WSI})$")

    for d in [0.2, 0.5, 1.0, 2.0]:
        vals = np.array([p_dfs_model1(30, d, c) - p_wsi_model1(30) for c in costs])
```

```
        axes[1].plot(costs, vals, label=fr"$¥delta={d:g}$")
    axes[1].axhline(0, lw=0.8, color="0.5")
    axes[1].set_title(r"B. Fixed budget $B=30$")
    axes[1].set_xlabel(r"Coarse observation cost per site $c$")
    axes[1].set_ylabel(r"$¥Delta P=P(¥mathrm{DFS})-P(¥mathrm{WSI})$")
    axes[1].legend(title="Coarse information", frameon=False)

    fig.suptitle(r"Model I: sensitivity to coarse observation cost ($¥delta=0.5$)", y=1.01)
    fig.tight_layout()
    save_figure(fig, "Figure_04_Model_I_cost_sensitivity")


#
============================================================
===
# Figure 5. Model II observation strategies (schematic)
#
============================================================
===
def figure5_model2_schematic():
    rows, cols = 6, 7
    target = (2, 4)
    # Fixed palette makes the schematic deterministic and visually heterogeneous.
    palette = [
        "#1f77b4", "#17becf", "#009e49", "#8e44ad", "#66615c", "#315aa6", "#e69f00",
        "#4aa3b8", "#d64f9b", "#ff7f0e", "#6f4e8b", "#735487", "#3b6b7a", "#1b9e91",
        "#5b6685", "#405b9a", "#86b817", "#5b8c25", "#d62728", "#9b2f57", "#3275bd",
        "#4f8094", "#914052", "#d36b75", "#777777", "#4c6578", "#74665f", "#ae6a6a",
        "#2474c6", "#1b9e77", "#5141a4", "#379399", "#aa3377", "#f0ad00", "#0b89a6",
        "#c43eab", "#00a6b2", "#3e7259", "#ae54a4", "#367184", "#7c4a33", "#bf379a",
    ]
    target_color = "#e41a1c"
    colors = list(palette)
    colors[target[0] * cols + target[1]] = target_color

    fig, axes = plt.subplots(1, 3, figsize=(14.5, 4.2))
```

```
titles = ["A    Model II ground truth", "B    WSI: fine trace", "C    DFS: coarse view → focus"]
for ax, title in zip(axes, titles):
    ax.set_title(title, loc="left", fontweight="bold")
    ax.set_xlim(-0.7, cols - 0.3)
    ax.set_ylim(-0.7, rows - 0.3)
    ax.set_aspect("equal")
    ax.set_xticks([]); ax.set_yticks([])

def draw_points(ax):
    for r in range(rows):
        for c in range(cols):
            ax.scatter(c, rows - 1 - r, s=48, c=colors[r * cols + c], zorder=3)

draw_points(axes[0])

# WSI trace, starting from lower left; 14 lower-row sites marked as fine-observed.
trace = []
for r in range(rows - 1, -1, -1):
    cc = range(cols) if (rows - 1 - r) % 2 == 0 else range(cols - 1, -1, -1)
    for c in cc:
        trace.append((r, c))
xy = [(c, rows - 1 - r) for r, c in trace]
axes[1].plot([a for a, b in xy], [b for a, b in xy], lw=1.0, alpha=0.75)
draw_points(axes[1])
for r, c in trace[:14]:
    axes[1].add_patch(Circle((c, rows - 1 - r), 0.20, fill=False, ec="0.2", lw=1.2, zorder=4))

# DFS selected candidates: schematic only.
draw_points(axes[2])
selected = {(0, 2), (0, 4), (1, 6), (2, 1), (2, 4), (3, 6), (4, 3), (5, 0), (5, 5)}
for r, c in selected:
    axes[2].add_patch(Circle((c, rows - 1 - r), 0.22, fill=False, ec="0.15", lw=1.2, zorder=4))
```

```
    fig.tight_layout()
    save_figure(fig, "Figure_05_Model_II_schematic")


#
============================================================
===
# Model II simulation engine used by Figures 6-9 and S1
#
============================================================
===
M2 = 50
DELTA2 = 1.0
KAPPA2 = 0.10
RHO2 = 0.90
SIGMA_ETA2 = 0.15
SIGMA_C2 = 0.70
SIGMA_LOCAL_C2 = 0.40
SIGMA_F2 = 0.20
SIGMA_LOCAL_F2 = 0.15
C_COARSE_2 = 0.05
C_FINE_2 = 1.0
B2 = np.arange(3, 51)
SB2 = np.linspace(0.0, 2.0, 41)


class Model2RandomBase:
    """Common random numbers shared across Model II strategies and parameter sweeps."""
    def __init__(self, n_trials=N2, seed=SEED):
        rng = np.random.default_rng(seed)
        self.n = n_trials
        self.u = rng.standard_normal((n_trials, M2))
        self.v = rng.standard_normal((n_trials, M2))
        self.eta = SIGMA_ETA2 * rng.standard_normal((n_trials, M2))
        self.eps_f = SIGMA_F2 * rng.standard_normal((n_trials, M2))
        self.eps_local_f = SIGMA_LOCAL_F2 * rng.standard_normal((n_trials, M2))
```

```
        self.eps_c = SIGMA_C2 * rng.standard_normal((n_trials, M2))
        self.eps_local_c = SIGMA_LOCAL_C2 * rng.standard_normal((n_trials, M2))
        self.priority = rng.random((n_trials, M2))
        self.target = np.zeros((1, M2))
        self.target[0, 0] = 1.0    # by exchangeability, target fixed at index 0
        tp = self.priority[:, 0]
        self.target_position = 1 + (self.priority < tp[:, None]).sum(axis=1)


MODEL2_BASE = Model2RandomBase()


def _first_higher_entry_position(score, base=MODEL2_BASE):
    higher = score > score[:, [0]]
    masked_priority = np.where(higher, base.priority, np.inf)
    min_priority = masked_priority.min(axis=1)
    out = np.full(base.n, M2 + 1, dtype=int)
    ok = np.isfinite(min_priority)
    if ok.any():
        out[ok] = 1 + (base.priority[ok] < min_priority[ok, None]).sum(axis=1)
    return out


def model2_condition(sigma_b, kappa=KAPPA2, rho=RHO2, coarse_cost=C_COARSE_2,
base=MODEL2_BASE):
    b = sigma_b * base.u
    b_local = sigma_b * (rho * base.u + np.sqrt(1.0 - rho**2) * base.v)

    x_fine = b + base.eta + DELTA2 * base.target
    x_coarse = b + base.eta + kappa * DELTA2 * base.target

    # Original WSI comparator: absolute redness.
    score_wsi_basic = x_fine + base.eps_f

    # Context-matched WSI comparator: same fine local-background subtraction as DFS.
    score_wsi_context = score_wsi_basic - (b_local + base.eps_local_f)
```

```
    # DFS coarse ranking score and fine score.
    score_dfs_coarse = (x_coarse + base.eps_c) - (b_local + base.eps_local_c)
    score_dfs_fine = score_wsi_context

    higher_basic_pos = _first_higher_entry_position(score_wsi_basic, base)
    higher_context_pos = _first_higher_entry_position(score_wsi_context, base)

    target_coarse_score = score_dfs_coarse[:, 0]
    target_coarse_rank = 1 + (score_dfs_coarse > target_coarse_score[:,
None]).sum(axis=1)

    higher_fine = score_dfs_fine > score_dfs_fine[:, [0]]
    best_coarse_among_higher_fine = np.where(higher_fine, score_dfs_coarse, -
np.inf).max(axis=1)
    first_higher_coarse_rank = np.full(base.n, M2 + 1, dtype=int)
    ok = np.isfinite(best_coarse_among_higher_fine)
    if ok.any():
        first_higher_coarse_rank[ok] = 1 + (
            score_dfs_coarse[ok] > best_coarse_among_higher_fine[ok, None]
        ).sum(axis=1)

    rows = []
    for B in B2:
        k_w = min(M2, int(math.floor(B / C_FINE_2)))
        m_d = max(0, min(M2, int(math.floor((B - M2 * coarse_cost) / C_FINE_2))))

        succ_basic = (base.target_position <= k_w) & (k_w < higher_basic_pos)
        succ_context = (base.target_position <= k_w) & (k_w < higher_context_pos)
        if m_d > 0:
            succ_dfs = (target_coarse_rank <= m_d) & (m_d < first_higher_coarse_rank)
        else:
            succ_dfs = np.zeros(base.n, dtype=bool)

        rows.append({
            "B": B,
```

```
            "P_WSI_basic": succ_basic.mean(),
            "P_WSI_context": succ_context.mean(),
            "P_DFS": succ_dfs.mean(),
            "Delta_context": succ_context.mean() - succ_basic.mean(),
            "Delta_selection": succ_dfs.mean() - succ_context.mean(),
            "Delta_total": succ_dfs.mean() - succ_basic.mean(),
        })
    return pd.DataFrame(rows)


def build_model2_main_grid():
    parts = []
    for sb in SB2:
        d = model2_condition(float(sb))
        d["sigma_b"] = sb
        parts.append(d)
    out = pd.concat(parts, ignore_index=True)
    out.to_csv(OUT / "Model_II_main_grid_20k.csv", index=False)
    return out


#
=============================================================
===
# Figure 6. Model II representative performance and crossover map
#
=============================================================
===
def figure6_model2_main(model2_main):
    fig, axes = plt.subplots(1, 2, figsize=(11.0, 4.5))

    # Panel A: representative sigma_b = 0.70
    d = model2_main[np.isclose(model2_main["sigma_b"], 0.70)]
    axes[0].plot(d["B"], d["P_WSI_basic"], label="WSI-AI")
    axes[0].plot(d["B"], d["P_DFS"], label="DFS-AI")
    axes[0].set_title(r"A. Representative budget dependence ($¥sigma_b=0.70$)")
```

```
    axes[0].set_xlabel(r"Total observation budget $B$")
    axes[0].set_ylabel("Detection probability")
    axes[0].set_xlim(3, 50)
    axes[0].set_ylim(0, 0.60)
    axes[0].legend(frameon=False)

    # Panel B: DeltaP = DFS - original WSI
    pivot          =          model2_main.pivot(index="sigma_b",          columns="B",
values="Delta_total").reindex(index=SB2, columns=B2)
    Z = pivot.to_numpy()
    vmax = max(abs(np.nanmin(Z)), abs(np.nanmax(Z)))
    mesh = axes[1].pcolormesh(
        B2, SB2, Z, shading="auto", cmap="RdBu_r",
        norm=TwoSlopeNorm(vmin=-vmax, vcenter=0, vmax=vmax),
    )
    axes[1].contour(B2, SB2, Z, levels=[0], colors="0.25", linewidths=1.0)
    axes[1].set_title("B. Budget × background heterogeneity")
    axes[1].set_xlabel(r"Total observation budget $B$")
    axes[1].set_ylabel(r"Background heterogeneity $¥sigma_b$")
    cb = fig.colorbar(mesh, ax=axes[1])
    cb.set_label(r"$¥Delta P=P(¥mathrm{DFS})-P(¥mathrm{WSI})$")

    fig.suptitle("Model II: representative performance and crossover map", y=1.01)
    fig.tight_layout()
    save_figure(fig, "Figure_06_Model_II_main")


#
============================================================
===
# Figure 7. Model II sensitivity to coarse signal visibility kappa
#
============================================================
===
def figure7_model2_kappa():
    fig, ax = plt.subplots(figsize=(7.0, 4.8))
```

```
    rows = []
    for kappa in [0.05, 0.10, 0.20, 0.40]:
        vals = []
        for sb in SB2:
            d = model2_condition(float(sb), kappa=kappa)
            r = d.loc[d["B"] == 30].iloc[0]
            dp = r["P_DFS"] - r["P_WSI_basic"]
            vals.append(dp)
            rows.append((kappa, sb, dp))
        ax.plot(SB2, vals, label=fr"$¥kappa={kappa:.2f}$")
    pd.DataFrame(rows, columns=["kappa", "sigma_b", "DeltaP"]).to_csv(OUT /
"Figure_07_data.csv", index=False)
    ax.axhline(0, lw=0.8, color="0.5")
    ax.set_title(r"Model II sensitivity to signal visibility in coarse observation ($B=30$)")
    ax.set_xlabel(r"Background heterogeneity $¥sigma_b$")
    ax.set_ylabel(r"$¥Delta P=P(¥mathrm{DFS})-P(¥mathrm{WSI})$")
    ax.legend(frameon=False)
    fig.tight_layout()
    save_figure(fig, "Figure_07_Model_II_kappa_sensitivity")


#
=================================================================
===
# Figure 8. Model II sensitivity to local-background correlation rho_l
#
=================================================================
===
def figure8_model2_rho():
    fig, ax = plt.subplots(figsize=(7.0, 4.8))
    rows = []
    for rho in [0.70, 0.80, 0.90, 0.97]:
        vals = []
        for sb in SB2:
            d = model2_condition(float(sb), rho=rho)
            r = d.loc[d["B"] == 30].iloc[0]
```

```
            dp = r["P_DFS"] - r["P_WSI_basic"]
            vals.append(dp)
            rows.append((rho, sb, dp))
        ax.plot(SB2, vals, label=fr"$¥rho_{{¥mathrm{{local}}}}={rho:.2f}$")
    pd.DataFrame(rows, columns=["rho_local", "sigma_b", "DeltaP"]).to_csv(OUT /
"Figure_08_data.csv", index=False)
    ax.axhline(0, lw=0.8, color="0.5")
    ax.set_title(r"Model II sensitivity to local background correlation ($B=30$)")
    ax.set_xlabel(r"Background heterogeneity $¥sigma_b$")
    ax.set_ylabel(r"$¥Delta P=P(¥mathrm{DFS})-P(¥mathrm{WSI})$")
    ax.legend(frameon=False)
    fig.tight_layout()
    save_figure(fig, "Figure_08_Model_II_rho_sensitivity")


#
===============================================================
===
# Figure 9. Model II sensitivity to coarse observation cost c
#
===============================================================
===
def figure9_model2_cost():
    fig, ax = plt.subplots(figsize=(7.0, 4.8))
    rows = []
    for cc in [0.02, 0.05, 0.10, 0.20]:
        vals = []
        for sb in SB2:
            d = model2_condition(float(sb), coarse_cost=cc)
            r = d.loc[d["B"] == 30].iloc[0]
            dp = r["P_DFS"] - r["P_WSI_basic"]
            vals.append(dp)
            rows.append((cc, sb, dp))
        ax.plot(SB2, vals, label=fr"$c={cc:.2f}$")
    pd.DataFrame(rows, columns=["coarse_cost", "sigma_b", "DeltaP"]).to_csv(OUT /
"Figure_09_data.csv", index=False)
```

```
    ax.axhline(0, lw=0.8, color="0.5")
    ax.set_title(r"Model II sensitivity to coarse observation cost ($B=30$)")
    ax.set_xlabel(r"Background heterogeneity $¥sigma_b$")
    ax.set_ylabel(r"$¥Delta P=P(¥mathrm{DFS})-P(¥mathrm{WSI})$")
    ax.legend(frameon=False)
    fig.tight_layout()
    save_figure(fig, "Figure_09_Model_II_cost_sensitivity")


#
============================================================
===
# Figure 10. Model III lesion structure schematic
#
============================================================
===
def figure10_model3_schematic():
    n = 7
    fig, axes = plt.subplots(1, 4, figsize=(14.5, 3.7))
    for ax, s in zip(axes, [1, 2, 3, 4]):
        ax.set_aspect("equal")
        ax.set_xlim(-0.7, n - 0.3)
        ax.set_ylim(-0.7, n - 0.3)
        ax.set_xticks([]); ax.set_yticks([])
        ax.set_title(f"Side length s = {s}¥nAffected lesion units: {s*s}", fontsize=9)
        for spine in ax.spines.values():
            spine.set_visible(False)

        start = (n - s) // 2
        lesion = {(r, c) for r in range(start, start + s) for c in range(start, start + s)}
        for r in range(n):
            for c in range(n):
                color = "tab:orange" if (r, c) in lesion else "tab:blue"
                ax.scatter(c, n - 1 - r, s=28, c=color)

        # Square outline around the lesion block.
```

```
            x0 = start - 0.45
            y0 = n - 1 - (start + s - 1) - 0.45
            ax.add_patch(Rectangle((x0, y0), s - 0.10, s - 0.10, fill=False, ec="tab:orange",
lw=1.5))

        fig.suptitle("Model III: one spatially continuous square lesion", y=1.02)
        fig.tight_layout()
        save_figure(fig, "Figure_10_Model_III_schematic")


#
============================================================
===
# Model III simulation engine used by Figures 11-13 and S2
#
============================================================
===
NGRID3 = 7
M3 = 49
DELTA3 = 0.5
C_COARSE_3 = 0.05
C_FINE_3 = 1.0
B3 = np.arange(3, 50)


def square_masks_model3(s):
    masks = []
    for r in range(NGRID3 - s + 1):
        for c in range(NGRID3 - s + 1):
            mask = np.zeros(M3, dtype=bool)
            for rr in range(r, r + s):
                mask[rr * NGRID3 + c: rr * NGRID3 + c + s] = True
            masks.append(mask)
    return np.asarray(masks)
```

```
def model3_rank_and_coverage(s, delta=DELTA3, coarse_cost=C_COARSE_3,
                             n_trials=N3, seed=None, batch_size=10_000):
    """
    Returns probabilities across B for one lesion side length.
    A deterministic seed is used for each s. For delta sensitivity, call with the same
    seed for all delta values so the comparison uses common random numbers.
    """
    if seed is None:
        seed = SEED + 1000 * s
    rng = np.random.default_rng(seed)
    L = s * s
    masks = square_masks_model3(s)
    Q = len(masks)
    mf = masks.astype(np.float32)

    max_m = max(0, min(Q, int(math.floor((B3.max() - M3 * coarse_cost) / (L *
C_FINE_3)))))
    exact_count = np.zeros(max_m + 1, dtype=np.int64)
    cov100_count = np.zeros(max_m + 1, dtype=np.int64)
    cov90_count = np.zeros(max_m + 1, dtype=np.int64)
    cov80_count = np.zeros(max_m + 1, dtype=np.int64)

    done = 0
    while done < n_trials:
        nb = min(batch_size, n_trials - done)
        true_idx = rng.integers(0, Q, size=nb)
        lesion = masks[true_idx]
        eps = rng.standard_normal((nb, M3)).astype(np.float32)
        x = eps + delta * lesion.astype(np.float32)
        field_score = (x @ mf.T) / np.sqrt(L)

        true_score = field_score[np.arange(nb), true_idx]
        true_rank = 1 + (field_score > true_score[:, None]).sum(axis=1)
        for m in range(1, max_m + 1):
            exact_count[m] += np.count_nonzero(true_rank <= m)
```

```
        if max_m > 0:
            order = np.argsort(-field_score, axis=1)[:, :max_m]
            union = np.zeros((nb, M3), dtype=bool)
            t90 = math.ceil(0.90 * L)
            t80 = math.ceil(0.80 * L)
            for t in range(max_m):
                union |= masks[order[:, t]]
                overlap = (union & lesion).sum(axis=1)
                m = t + 1
                cov100_count[m] += np.count_nonzero(overlap >= L)
                cov90_count[m] += np.count_nonzero(overlap >= t90)
                cov80_count[m] += np.count_nonzero(overlap >= t80)
        done += nb

    rows = []
    for B in B3:
        k_w = min(M3, int(math.floor(B / C_FINE_3)))
        m_d = max(0, min(Q, int(math.floor((B - M3 * coarse_cost) / (L * C_FINE_3)))))
        k_field = max(0, min(Q, int(math.floor(B / (L * C_FINE_3)))))

        p_discovery = hypergeom.sf(0, M3, L, k_w)
        p_complete = math.comb(k_w, L) / math.comb(M3, L) if k_w >= L else 0.0
        p_field_exact = k_field / Q

        rows.append({
            "s": s, "L": L, "Q": Q, "B": B,
            "P_WSI_discovery": p_discovery,
            "P_WSI_complete": p_complete,
            "P_field_exact": p_field_exact,
            "P_DFS_exact": exact_count[m_d] / n_trials,
            "P_DFS_union100": cov100_count[m_d] / n_trials,
            "P_WSI_cov90": hypergeom.sf(math.ceil(0.90 * L) - 1, M3, L, k_w),
            "P_DFS_cov90": cov90_count[m_d] / n_trials,
            "P_WSI_cov80": hypergeom.sf(math.ceil(0.80 * L) - 1, M3, L, k_w),
            "P_DFS_cov80": cov80_count[m_d] / n_trials,
        })
```

```
    return pd.DataFrame(rows)


def build_model3_main():
    out = pd.concat([model3_rank_and_coverage(s) for s in [1, 2, 3, 4]],
ignore_index=True)
    out.to_csv(OUT / "Model_III_main_and_coverage_200k.csv", index=False)
    return out


#
==========================================================
===
# Figure 11. Model III discovery and complete characterization
#
==========================================================
===
def figure11_model3_main(model3_main):
    fig, axes = plt.subplots(2, 2, figsize=(10.0, 7.3), sharex=True, sharey=True)
    for ax, s in zip(axes.ravel(), [1, 2, 3, 4]):
        d = model3_main[model3_main["s"] == s]
        L = s * s
        ax.plot(d["B"], d["P_WSI_discovery"], label="WSI-AI: detect ≥1 lesion unit")
        ax.plot(d["B"], d["P_WSI_complete"], label="WSI-AI: 100% lesion-unit coverage")
        ax.plot(d["B"], d["P_DFS_exact"], label="DFS-AI: exact lesion-field selection")
        ax.set_title(fr"$s={s}$ ($L={L}$)")
        ax.set_xlabel(r"Total observation budget $B$")
        ax.set_ylabel("Success probability")
        ax.set_xlim(3, 49); ax.set_ylim(0, 1.02)
    handles, labels = axes[0, 0].get_legend_handles_labels()
    fig.legend(handles, labels, loc="lower center", ncol=1, frameon=False,
bbox_to_anchor=(0.5, -0.01))
    fig.suptitle("Model III: discovery and complete characterization under finite resources",
y=1.01)
    fig.tight_layout(rect=[0, 0.10, 1, 1])
    save_figure(fig, "Figure_11_Model_III_discovery_complete")
```

```
#
===========================================================
===
# Figure 12. Model III DFS advantage for complete characterization
#
===========================================================
===
def figure12_model3_advantage(model3_main):
    fig, ax = plt.subplots(figsize=(7.5, 4.8))
    for s in [1, 2, 3, 4]:
        d = model3_main[model3_main["s"] == s]
        dp = d["P_DFS_exact"].to_numpy() - d["P_WSI_complete"].to_numpy()
        ax.plot(d["B"], dp, label=fr"$s={s}$")
    ax.axhline(0, lw=0.8, color="0.5")
    ax.set_title("Model III: DFS advantage for complete lesion characterization")
    ax.set_xlabel(r"Total observation budget $B$")
    ax.set_ylabel(r"$¥Delta                                  P=P(¥mathrm{DFS,exact})-
P(¥mathrm{WSI,100¥%¥ coverage})$")
    ax.legend(frameon=False)
    fig.tight_layout()
    save_figure(fig, "Figure_12_Model_III_DFS_advantage")


#
===========================================================
===
# Figure 13. Model III sensitivity to delta and coarse cost
#
===========================================================
===
def figure13_model3_sensitivity():
    deltas = np.arange(0.0, 2.01, 0.25)
    costs = np.arange(0.0, 0.301, 0.05)
    B_fixed = 30
```

```
fig, axes = plt.subplots(1, 2, figsize=(11.0, 4.4))
sens_rows = []

# Panel A: varying delta. Same seed for all delta values at a given s.
for s in [1, 2, 3, 4]:
    vals = []
    for dlt in deltas:
        sim = model3_rank_and_coverage(s, delta=float(dlt), coarse_cost=C_COARSE_3,
                                       seed=SEED + 1000 * s)
        r = sim.loc[sim["B"] == B_fixed].iloc[0]
        dp = r["P_DFS_exact"] - r["P_WSI_complete"]
        vals.append(dp)
        sens_rows.append(("delta", s, dlt, dp))
    axes[0].plot(deltas, vals, marker="o", ms=3, label=fr"$s={s}$")
axes[0].axhline(0, lw=0.8, color="0.5")
axes[0].set_title("A. Sensitivity to coarse information")
axes[0].set_xlabel(r"Coarse view discriminability per lesion unit $¥delta$")
axes[0].set_ylabel(r"$¥Delta P$ at $B=30$")
axes[0].legend(frameon=False)

# Panel B: varying coarse cost. Reuse the same underlying baseline rank process.
# We call the simulator separately here so that the code mirrors the stated model directly.
for s in [1, 2, 3, 4]:
    vals = []
    for cc in costs:
        sim = model3_rank_and_coverage(s, delta=DELTA3, coarse_cost=float(cc),
                                       seed=SEED + 1000 * s)
        r = sim.loc[sim["B"] == B_fixed].iloc[0]
        dp = r["P_DFS_exact"] - r["P_WSI_complete"]
        vals.append(dp)
        sens_rows.append(("coarse_cost", s, cc, dp))
    axes[1].plot(costs, vals, marker="o", ms=3, label=fr"$s={s}$")
axes[1].axhline(0, lw=0.8, color="0.5")
axes[1].set_title("B. Sensitivity to coarse observation cost")
```

```
    axes[1].set_xlabel(r"Coarse observation cost $c$")
    axes[1].set_ylabel(r"$¥Delta P$ at $B=30$")
    axes[1].legend(frameon=False)

    pd.DataFrame(sens_rows, columns=["parameter", "s", "value", "DeltaP"]).to_csv(
        OUT / "Figure_13_data.csv", index=False
    )
    fig.suptitle("Model III: sensitivity analysis", y=1.01)
    fig.tight_layout()
    save_figure(fig, "Figure_13_Model_III_sensitivity")


#
==============================================================
===
# Supplementary Figure S1. Model II context-matched WSI control
#
==============================================================
===
def supplementary_figure_s1(model2_main):
    fig, axes = plt.subplots(1, 3, figsize=(15.0, 4.4))

    d30 = model2_main[model2_main["B"] == 30]
    axes[0].plot(d30["sigma_b"], d30["P_WSI_basic"], label="WSI-AI: basic")
    axes[0].plot(d30["sigma_b"],    d30["P_WSI_context"],    label="WSI-AI:    context
matched")
    axes[0].plot(d30["sigma_b"], d30["P_DFS"], label="DFS-AI")
    axes[0].set_title(r"A. Matched comparison at $B=30$")
    axes[0].set_xlabel(r"Background heterogeneity $¥sigma_b$")
    axes[0].set_ylabel("Detection probability")
    axes[0].legend(frameon=False)

    def hm(ax, column, title, cbar_label):
        pivot        =        model2_main.pivot(index="sigma_b",        columns="B",
values=column).reindex(index=SB2, columns=B2)
        Z = pivot.to_numpy()
```

```
            vmax = max(abs(np.nanmin(Z)), abs(np.nanmax(Z)))
            mesh = ax.pcolormesh(B2, SB2, Z, shading="auto", cmap="RdBu_r",
                                             norm=TwoSlopeNorm(vmin=-vmax,              vcenter=0,
vmax=vmax))
            ax.contour(B2, SB2, Z, levels=[0], colors="0.25", linewidths=0.9)
            ax.set_xlabel(r"Total observation budget $B$")
            ax.set_ylabel(r"Background heterogeneity $¥sigma_b$")
            ax.set_title(title)
            cb = fig.colorbar(mesh, ax=ax)
            cb.set_label(cbar_label)

      hm(axes[1], "Delta_context", "B. Context effect",
          r"$P(¥mathrm{WSI,context})-P(¥mathrm{WSI,basic})$")
      hm(axes[2], "Delta_selection", "C. Selection effect after context matching",
          r"$P(¥mathrm{DFS})-P(¥mathrm{WSI,context})$")

      fig.suptitle("Supplementary Figure S1. Model II matched control decomposition",
y=1.02)
      fig.tight_layout()
      save_figure(fig, "Supplementary_Figure_S1_Model_II_context_matched")


#
============================================================
===
# Supplementary Figure S2. Model III field control and coverage sensitivity
#
============================================================
===
def supplementary_figure_s2(model3_main):
      fig, axes = plt.subplots(2, 3, figsize=(13.5, 7.5))
      axes = axes.ravel()

      specs = [
            ("A. Structured field allocation effect",
             lambda d: d["P_field_exact"] - d["P_WSI_complete"],
```

```
            r"$P(¥mathrm{random¥ field})-P(¥mathrm{WSI,100¥%})$"),
        ("B. Coarse selection beyond field matching",
            lambda d: d["P_DFS_exact"] - d["P_field_exact"],
            r"$P(¥mathrm{DFS,exact})-P(¥mathrm{random¥ field})$"),
        ("C. Symmetric 100% lesion-unit coverage",
            lambda d: d["P_DFS_union100"] - d["P_WSI_complete"],
            r"$P(¥mathrm{DFS,100¥%})-P(¥mathrm{WSI,100¥%})$"),
        ("D. Robustness: ≥90% lesion coverage",
            lambda d: d["P_DFS_cov90"] - d["P_WSI_cov90"],
            r"$P(¥mathrm{DFS,90¥%})-P(¥mathrm{WSI,90¥%})$"),
        ("E. Robustness: ≥80% lesion coverage",
            lambda d: d["P_DFS_cov80"] - d["P_WSI_cov80"],
            r"$P(¥mathrm{DFS,80¥%})-P(¥mathrm{WSI,80¥%})$"),
    ]

    for ax, (title, func, ylabel) in zip(axes[:5], specs):
        for s in [1, 2, 3, 4]:
            d = model3_main[model3_main["s"] == s]
            ax.plot(d["B"], func(d), label=fr"$s={s}$")
        ax.axhline(0, lw=0.8, color="0.5")
        ax.set_title(title, fontsize=9)
        ax.set_xlabel(r"Total observation budget $B$")
        ax.set_ylabel(ylabel, fontsize=8)
        ax.legend(frameon=False, fontsize=7)
    axes[5].axis("off")

    fig.suptitle("Supplementary Figure S2. Model III field control and coverage sensitivity",
y=1.01)
    fig.tight_layout()
    save_figure(fig, "Supplementary_Figure_S2_Model_III_field_control")


#
==============================================================
===
# Main execution: figures are generated in manuscript order
```

```
#
==========================================================
===
def main():
    print(f"QUICK={QUICK}; Model II trials={N2:,}; Model III trials={N3:,}")

    print("Generating Figure 1 ...")
    figure1_model1_schematic()

    print("Generating Figure 2 ...")
    figure2_model1_budget()

    print("Generating Figure 3 ...")
    figure3_model1_delta_map()

    print("Generating Figure 4 ...")
    figure4_model1_cost_sensitivity()

    print("Generating Figure 5 ...")
    figure5_model2_schematic()

    print("Computing Model II grid ...")
    model2_main = build_model2_main_grid()

    print("Generating Figure 6 ...")
    figure6_model2_main(model2_main)

    print("Generating Figure 7 ...")
    figure7_model2_kappa()

    print("Generating Figure 8 ...")
    figure8_model2_rho()

    print("Generating Figure 9 ...")
    figure9_model2_cost()
```

```
    print("Generating Figure 10 ...")
    figure10_model3_schematic()

    print("Computing Model III main simulation ...")
    model3_main = build_model3_main()

    print("Generating Figure 11 ...")
    figure11_model3_main(model3_main)

    print("Generating Figure 12 ...")
    figure12_model3_advantage(model3_main)

    print("Generating Figure 13 ...")
    figure13_model3_sensitivity()

    print("Generating Supplementary Figure S1 ...")
    supplementary_figure_s1(model2_main)

    print("Generating Supplementary Figure S2 ...")
    supplementary_figure_s2(model3_main)

    print("Done. Output directory:", OUT.resolve())


if __name__ == "__main__":
    main()
```